\documentclass[onecolumn,preprintnumbers,amsmath,amssymb]{revtex4-2}
\usepackage[utf8]{inputenc}
\usepackage{enumitem}
\usepackage{epsfig}
\usepackage{graphicx}
\usepackage{dcolumn}
\usepackage{bm}
\usepackage{amsmath,amsfonts}
\usepackage{amssymb}
\usepackage{physics}
\usepackage{color}
\usepackage{xcolor}
\usepackage[colorlinks,linkcolor=blue,anchorcolor=blue,urlcolor=blue,citecolor=blue]{hyperref}
\usepackage{siunitx,booktabs}
\usepackage{epstopdf}
\usepackage{dsfont}
\begin{document}
	\title{A fully passive transmitter for decoy-state quantum key distribution}
	\author{Víctor Zapatero$^{1,2,3}$}
	\email{vzapatero@com.uvigo.es}
	\author{Wenyuan Wang$^{4}$}
	\author{Marcos Curty$^{1,2,3}$}
\affiliation{$^1$Vigo Quantum Communication Center, University of Vigo, Vigo E-36310, Spain}
\affiliation{$^2$Escuela de Ingeniería de Telecomunicación, Department of Signal Theory and Communications, University of Vigo, Vigo E-36310, Spain}
\affiliation{$^3$AtlanTTic Research Center, University of Vigo, Vigo E-36310, Spain}
\affiliation{$^{4}$Department of Physics, University of Hong Kong, Pokfulam Road, Hong Kong}
\begin{abstract}
	A passive quantum key distribution (QKD) transmitter generates the quantum states prescribed by a QKD protocol at random, combining a fixed quantum mechanism and a post-selection step. By circumventing the use of active optical modulators externally driven by random number generators, passive QKD transmitters offer immunity to modulator side channels and potentially enable higher frequencies of operation. Recently, the first linear optics setup suitable for passive decoy-state QKD has been proposed. In this work, we simplify the prototype and adopt sharply different approaches for BB84 polarization encoding and decoy-state parameter estimation. In particular, our scheme avoids a probabilistic post-selection step that is central to the former proposal. On top of it, we elaborate a simple and tight custom-made security analysis.
\end{abstract}
\maketitle
\section{Introduction}\label{Introduction}
Quantum key distribution (QKD) allows for information-theoretically secure key exchange between distant parties through an insecure channel~\cite{PortmannRenner,Curty}. This possibility, which is inaccessible from the point of view of classical communications, makes QKD a promising candidate for long-term communication security. Nowadays, QKD represents one of the most mature applications of quantum information science, and it is expected to become a prolific industry in the years to come. Notwithstanding, the security of real QKD implementations is not fully established yet, due to the difficulty of experimentally guaranteeing that the QKD devices stick to the assumptions and models presumed in the security proofs~\cite{Feihu}.

A particularly controversial assumption of most QKD security analyses is that no information leakage occurs through the boundaries of Alice's and Bob's labs. This premise opens the door for the so-called Trojan horse attacks (THAs)~\cite{Vakhitov,Gisin,Jain1,Jain2,Sajeed}, where an adversary injects bright light pulses into a QKD transmitter/receiver and then measures the back-reflected light, aiming to extract information about the setting choices. Notably, a possible solution to deal with information leakage in the QKD transmitter consists of trying to upper bound Eve's accessible information gain and account for it in the estimation of the secret key length~\cite{Lucamarini,Tamaki,Weilong,Navs}. Note, however, that this approach relies on modelling the information leakage to a certain extent, and it requires to add significant optical isolation to prevent a severe drop of the secret key rate and the achievable distance.

\begin{figure}[!htbp]
	\centering 
	\includegraphics[width=12.3cm,height=2.4cm]{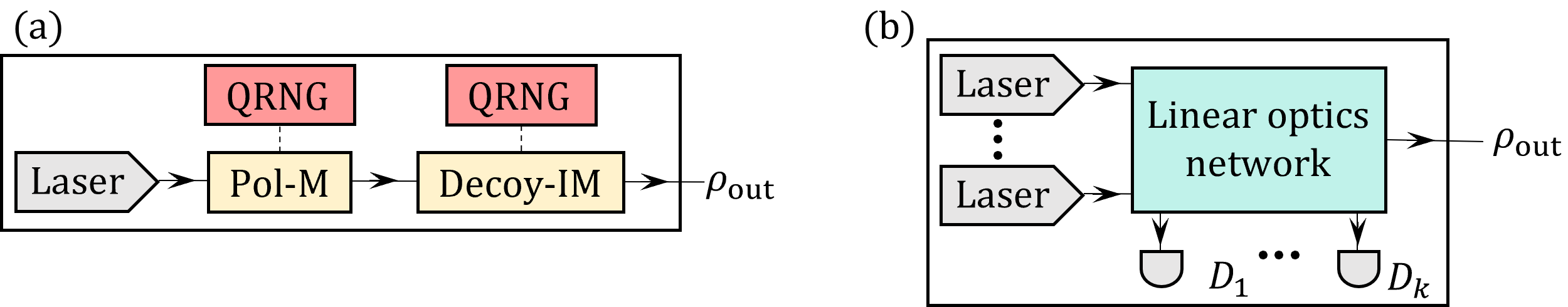}\\
	\caption{Schematic depiction of (a) active and (b) passive decoy-state QKD transmission. In the active case, a polarization modulator (Pol-M) and an intensity modulator (Decoy-IM) are driven by respective quantum random number generators (QRNGs), responsible for implementing the protocol settings. In the passive case, one or more laser diodes generate pulses that are sent through a linear optics network. Depending on the detection outcomes observed in the photodetectors ($\{D_{j}\}$), different signal states are actually generated, avoiding the use of active modulators or QRNGs.}
	\label{active_vs_passive}
\end{figure}

On the contrary, an alternative solution that might rule out THAs once and for all is to consider a fully passive (rather than active) QKD transmitter, as illustrated in Fig.~\ref{active_vs_passive}.

In a passive transmitter (PT), the protocol states are generated at random using inherent quantum randomness of the device, in so avoiding the use of quantum random number generators (QRNGs) to actively modulate the protocol settings (\textit{e.g.} intensity, phase or polarization), or the use of auxiliary optical modulators of any kind. Indeed, the advantages of passive encoding go beyond the obvious security upgrade. To be precise, the fact that a PT avoids using externally driven elements makes it very desirable to operate QKD systems at high transmission rates, and to reduce the complexity (and thus the cost) of practical QKD implementations~\cite{Liu}. Indeed, the possibility to reach an enhanced bandwidth by suppressing active modulation has already been realized in~\cite{Paraiso}, where the construction a modulator-free (although active) QKD transmission chip is reported.

However, these advantages come at the price of decreasing the key generation rate for two main reasons. On the one hand, in a PT, additional sifting is required to discard those rounds where the randomly generated settings do not lie in certain acceptance regions. On the other hand, the finite size of these acceptance regions is an inherent source of noise not present in the active case.

Various proposals exist for passive decoy-state generation with parametric down-conversion sources~\cite{PDC_1,PDC_2,PDC_3,PDC_4,PDC_5,PDC_6}, or using coherent light~\cite{WCP_1,WCP_2,WCP_3,WCP_4}. Notably, both alternatives have been demonstrated experimentally~\cite{experiment_1,experiment_2,experiment_3,experiment_4,experiment_5,experiment_6}. In parallel, a simple alternative to passively generate random photon polarizations in a plane was reported in~\cite{passive_BB84}, suitable for a passive implementation of the BB84 protocol~\cite{BB84}. What is more, a PT preparing decoy-state BB84 signals with coherent pulses was introduced in~\cite{nonlinear}. Nevertheless, this latter proposal relies on a non-linear optical effect called sum-frequency generation~\cite{nonlinear_2}, which reduces its practicality.

In short, a simple setup simultaneously generating random decoy states and random photon polarizations in a fully passive way remained elusive for more than a decade. Recently though, a linear-optics-based PT of this kind has been proposed in~\cite{Mike}, combining the ideas of~\cite{WCP_2} and~\cite{passive_BB84}. Specifically, in~\cite{Mike}, a passive decoy-state BB84 protocol is considered, using a single intensity for the key generation basis and three different intensities for the parameter estimation basis. In the present work, we devise a simplified architecture for the prototype presented there and consider the standard decoy-state BB84 protocol instead, with three common intensity settings per basis. For symmetry reasons, our protocol uses both bases for key generation.

Remarkably as well, to avoid the assessment of the security analysis with mixed polarization states, the proposal in~\cite{Mike} relies on a facilitating assumption. Namely, that the actual protocol can be reduced to an ideal one where perfect Bell pairs are prepared for the single photon events and the effect of the mixed polarizations is incorporated a posteriori, adding post-processing noise at Alice's local measurements. Here, we circumvent this assumption by elaborating a custom-made security analysis for the mixed single-photon states.

On top of it, the decoy-state method in~\cite{Mike} relies on an auxiliary post-selection probability to decouple the intensity and the polarization of the output Fock states of the PT in the parameter estimation basis. This step, which entails an undesired discard of raw data, requires a QRNG and might be cumbersome to implement in practice. In this work, we avoid this extra sifting by tackling the decoy-state parameter estimation with the intensity-setting-dependent Fock states directly. 

The structure of the paper goes as follows. We present the PT and its mixed output state in Sec.~\ref{Characterization}. In Sec.~\ref{post-selection} we describe a simple approach to post-select decoy-state BB84 acceptance regions, together with the quantum states that arise from this post-selection. Sec.~\ref{decoy-state} is dedicated to explain our decoy-state parameter estimation method, and in Sec.~\ref{PHER} we derive the single-photon phase-error rate of the problem at hand. Coming next, in Sec.~\ref{Performance} we evaluate the rate-distance performance of our passive QKD scheme and compare it to the performance achieved in the active setting. Finally, Sec.~\ref{Discussion} provides a summarizing discussion, and a series of appendices are included at the end of the paper to ensure the reproducibility of the results.
\section{A passive QKD transmitter}\label{Characterization}
\begin{figure}[!htbp]
	\centering 
	\includegraphics[width=9.4cm,height=4.2cm]{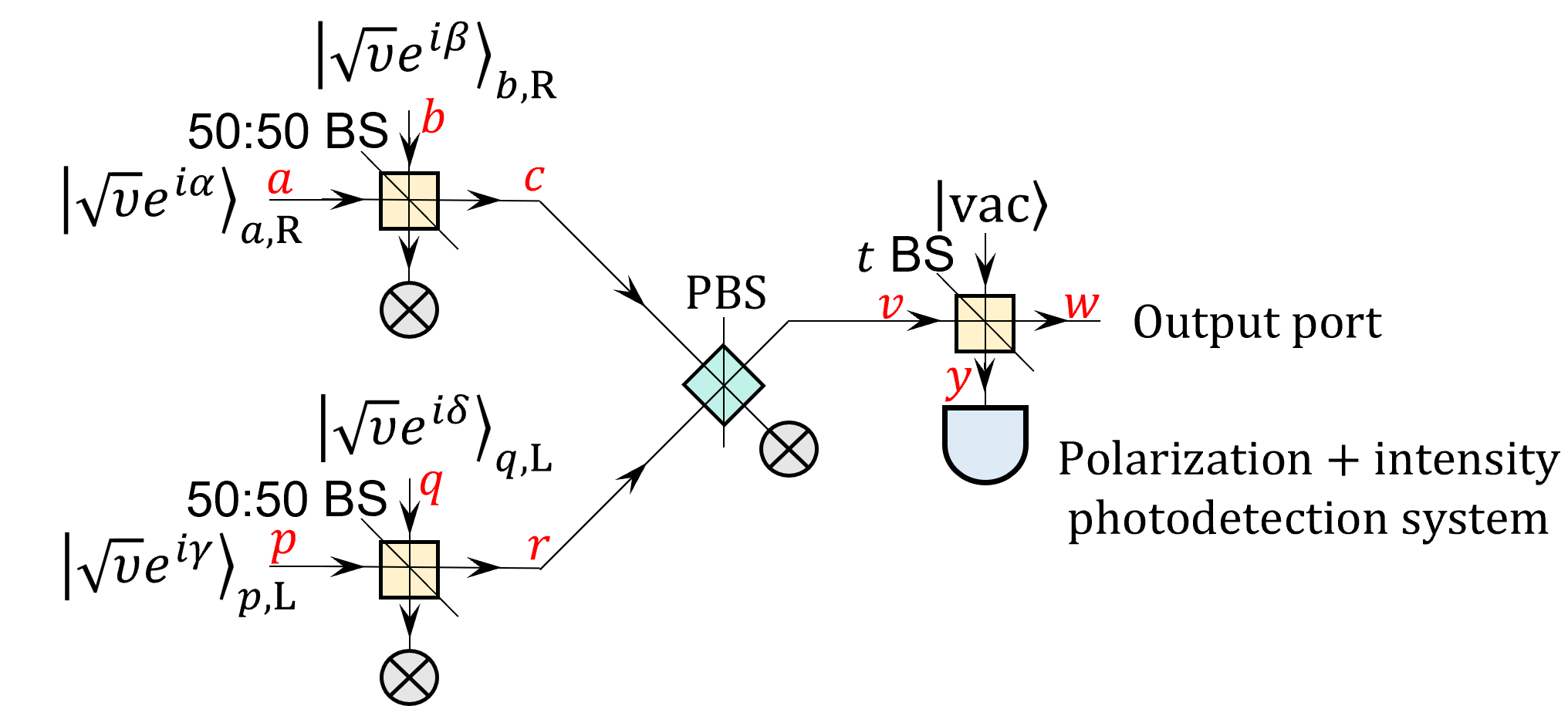}\\
	\caption{Architecture of the PT. All input coherent states in the figure have a common large intensity $\nu$ and independent random phases $\alpha$, $\beta$, $\gamma$ and $\delta$. Red color is used to denote the relevant spatial modes, while unused modes are tagged by the symbol ``$\otimes$". The interference occurring in the 50:50 beamsplitter (BS) of the top (bottom) arm yields a coherent state with a random intensity dependent on the phase difference $\beta-\alpha$ ($\delta-\gamma$), while the polarization ---common to both input states--- is preserved. Crucially, the polarization of the top arm (right-handed) is orthogonal to that of the bottom arm (left-handed). When combined in the polarizing BS (PBS), these orthogonally polarized coherent pulses generate a final coherent state with random intensity and random polarization in the RL Bloch sphere, coupled to each other. Lastly, this state enters a BS with transmittance $t\ll{1}$, where the intensity is attenuated to the single-photon level. Importantly, while the transmitted signal goes to the quantum channel, the reflected signal reaches a photo-detection system that accurately measures its polarization and intensity for post-selection purposes.}
	\label{schematic}
\end{figure}

The PT we propose is depicted in Fig.~\ref{schematic}. In the figure, we use the notation $\ket{\tau}_{a,\rm R(L)}$ to denote a right-handed (left-handed) circularly polarized weak coherent pulse (WCP) in the spatial mode $a$ with complex amplitude $\tau\in\mathbb{C}$. Namely, $\ket{\tau}_{a,\rm R(L)}=\exp\bigl\{\tau{}a_{\rm R(L)}^{\dagger}-\tau^{*}{}a_{\rm R(L)}\bigr\}\ket{\rm vac}$, where $\ket{\rm vac}$ is the vacuum state and $\{a^{\dagger}_{\rm R(L)},a_{\rm R(L)}\}$ denote the creation/annihilation operators of a right-handed (left-handed) circular polarization state in spatial mode $a$. Notably, $a^{\dagger}_{\rm R}\ket{\rm vac}$ and $a^{\dagger}_{\rm L}\ket{\rm vac}$ shall be viewed as the north and the south pole of a Bloch sphere throughout this work, which we shall refer to as ``the RL Bloch sphere". The selection of the circularly polarized single-photon states as the poles of the sphere is such that the equator plane of the sphere contains all possible linearly polarized single-photon states, the specific orientation being determined by the azimuthal angle. In particular, the creation operator associated to an arbitrary polarization $(\theta,\phi)$ in the RL Bloch sphere reads
\begin{equation}\label{creation_operator}
a^{\dagger}_{\theta,\phi}=\cos\left(\frac{\theta}{2}\right)a^{\dagger}_{\rm R}+e^{i\phi}\sin\left(\frac{\theta}{2}\right)a^{\dagger}_{\rm L},
\end{equation}
where $\theta$ ($\phi$) stands for the polar (azimuthal) angle of the sphere. Also, a WCP in spatial mode $a$, with amplitude $\tau$ and polarization specified by $(\theta,\phi)$ shall be denoted as $\ket{\tau}_{a,\theta,\phi}$. This said, let us present the output state of the PT.

Instead of referring to the independent and identically distributed phases $\alpha$, $\beta$, $\gamma$ and $\delta$ of Fig.~\ref{schematic}, the output state of the PT is better described in terms of the parameters $\alpha$, $\delta_{1}=\beta-\alpha$, $\delta_{2}=\gamma-\beta$ and $\delta_{3}=\delta-\gamma$, the last three phase differences being uniformly random and independent too. Particularly, for specific values of $\alpha$, $\delta_{1}$, $\delta_{2}$ and $\delta_{3}$, the output state at mode $w$ in Fig.~\ref{schematic} reads
\begin{equation}\label{not_randomised}
\ket{\Psi}_{w}=\ket{\sqrt{I(\delta_{1},\delta_{3})}\ e^{i\psi(\alpha,\delta_{1})}}_{w,\theta(\delta_{1},\delta_{3}),\phi(\delta_{1},\delta_{2},\delta_{3})},
\end{equation}
where the quantities $I(\delta_{1},\delta_{3})$, $\psi(\alpha,\delta_{1})$, $\theta(\delta_{1},\delta_{3})$ and $\phi(\delta_{1},\delta_{2},\delta_{3})$ are given by
\begin{eqnarray}\label{variables}
&&I(\delta_{1},\delta_{3})=2\nu{}t\left[\sin^{2}\left(\frac{\delta_{1}}{2}\right)+\sin^{2}\left(\frac{\delta_{3}}{2}\right)\right]\hspace{.2cm}\mathrm{(intensity)},\nonumber \\
&&\psi(\alpha,\delta_{1})=\alpha+\frac{\delta_{1}}{2}-\mathrm{sgn}(\delta_{1})\frac{\pi}{2}\hspace{.2cm}\mathrm{(phase)},\nonumber \\
&&\theta(\delta_{1},\delta_{3})=2\arctan\left[\sin\left(\frac{\delta_{3}}{2}\right)\biggl/\sin\left(\frac{\delta_{1}}{2}\right)\right]\hspace{.2cm}\mathrm{(polar\ angle\ in\ the\ RL\ Bloch\ sphere)},\nonumber \\
&&\phi(\delta_{1},\delta_{2},\delta_{3})=\delta_{2}+\frac{\delta_{1}+\delta_{3}}{2}-\frac{\pi}{2}\Bigl[\mathrm{sgn}(\delta_{3})-\mathrm{sgn}(\delta_{1})\Bigr]\hspace{.2cm}\mathrm{(azimuthal\ angle\ in\ the\ RL\ Bloch\ sphere)}.
\end{eqnarray}
This is proven in Appendix~\ref{quantum_optics} using standard linear quantum optics.

If we now assume perfect phase-randomisation for the input coherent states, the mixed output state of the transmitter is
\begin{equation}\label{averaging}
\sigma_{w}=\frac{1}{(2\pi)^{4}}\int_{0}^{2\pi}\int_{0}^{2\pi}\int_{0}^{2\pi}\int_{0}^{2\pi}d\alpha{}d\delta_{1}d\delta_{2}d\delta_{3}\ketbra{\Psi}{\Psi}_{w}.
\end{equation}
The integral in $\alpha$ is direct (the purpose of changing variables from $\beta$, $\gamma$ and $\delta$ to $\delta_{1}$, $\delta_{2}$ and $\delta_{3}$ is to enable this immediate integration in $\alpha$) and yields a phase-randomised WCP,
\begin{equation}\label{PRWCP}
\frac{1}{2\pi}\int_{0}^{2\pi}d\alpha\ketbra{\Psi}{\Psi}_{w}=\sum_{n=0}^{\infty}\frac{e^{-I}{I}^{n}}{n!}\ketbra{n}{n}_{\theta,\phi},
\end{equation}
with the $n$-photon Fock states $\ket{n}_{\theta,\phi}$ given by
\begin{equation}\label{fock_theta_phi}
\ket{n}_{\theta,\phi}=\frac{\left(w^{\dagger}_{\theta,\phi}\right)^{n}}{\sqrt{n!}}\ket{\rm vac}.
\end{equation}
Note that we have omitted the dependence on $\delta_{1}$, $\delta_{2}$ and $\delta_{3}$ in Eq.~(\ref{PRWCP}) for readability. Also, the spatial mode $w$ is not explicitly indicated any more.

The phase differences $\delta_{1}$, $\delta_{2}$ and $\delta_{3}$ uniquely determine the relevant output variables $\phi$, $\theta$ and $I$, and it is desirable to describe the output state of the transmitter (given in Eq.~(\ref{averaging})) in terms of the latter instead. The joint probability density function $f_{\boldsymbol{\phi},\boldsymbol{\theta},\boldsymbol{I}}(\phi,\theta,I)$ is obtained in Appendix~\ref{distribution} and factors as
\begin{equation}\label{PDF_3}
f_{\boldsymbol{\phi},\boldsymbol{\theta},\boldsymbol{I}}(\phi,\theta,I)=f_{\boldsymbol{\phi}}(\phi)\times{}f_{\boldsymbol{\theta},\boldsymbol{I}}(\theta,I),
\end{equation}
where
\begin{equation}\label{PDF_2}
f_{\boldsymbol{\phi}}(\phi)=\frac{1}{2\pi}\hspace{.2cm}\mathrm{and}\hspace{.2cm}f_{\boldsymbol{\theta},\boldsymbol{I}}(\theta,I)=\frac{1}{2\nu{}t{}\pi^{2}\sqrt{1-\frac{I}{2\nu{}t}\displaystyle{\cos^{2}\left(\frac{\theta}{2}\right)}}\sqrt{1-\frac{I}{2\nu{}t}\displaystyle{\sin^{2}\left(\frac{\theta}{2}\right)}}},
\end{equation}
for $\phi\in(-\pi,\pi]$, $\theta\in[0,\pi]$ and $I\in\left[0,I_{\mathrm{max},\theta}\right)$, $I_{\mathrm{max},\theta}$ being defined as $I_{\mathrm{max},\theta}=\min\left\{2\nu{}t/\cos^{2}\left(\theta/2\right),2\nu{}t/\sin^{2}\left(\theta/2\right)\right\}$.\\

Notably, the above distribution exhibits azimuthal symmetry and is peaked towards the equator plane of the RL Bloch sphere, given by $\theta=\pi/2$. This makes the proposed architecture convenient for a passive decoy-state BB84 protocol using ``equator-plane BB84 states". In the next section, we elaborate on this idea.

\section{Post-selection of BB84 acceptance regions}\label{post-selection}
\begin{figure}[!htbp]
	\centering 
	\includegraphics[width=10cm,height=4.8cm]{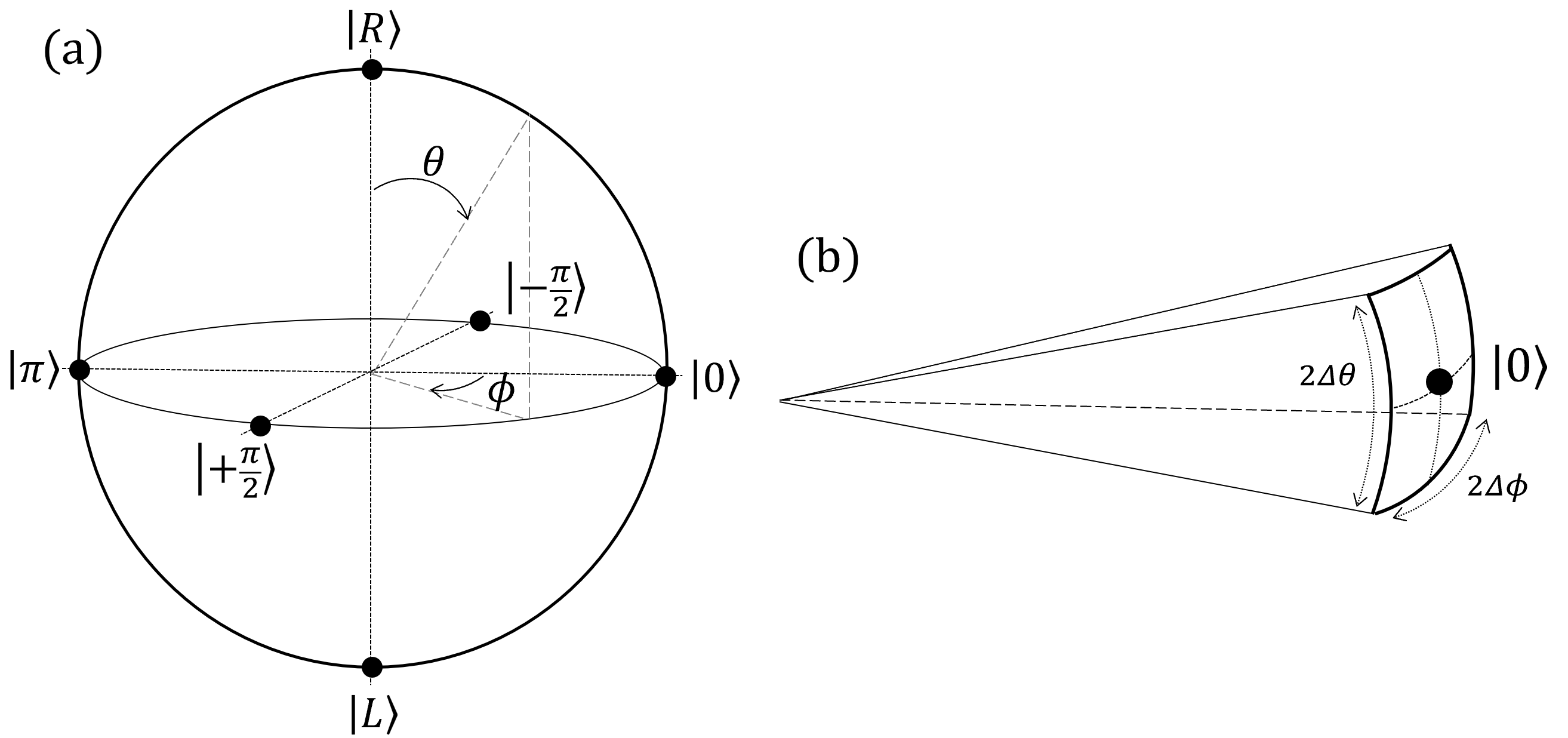}\\
	\caption{Post-selection of BB84 acceptance regions in the RL Bloch sphere. The sphere is illustrated in figure (a). $\theta$ ($\phi$) provides the polar (azimuthal) angle, $\ket{R}=a^{\dagger}_{\rm R}\ket{\rm vac}$ ($\ket{L}=a^{\dagger}_{\rm L}\ket{\rm vac}$) denotes a right-handed (left-handed) circularly polarized single-photon state, and $\left\{\ket{x}=\left(\ket{\rm R}+e^{ix}\ket{\rm L}\right)/\sqrt{2}\right\}$ with $x\in\{0,\pi,\pi/2,-\pi/2\}$ defines a set of four linearly polarized single-photon states evenly spaced in the equator line, suitable for an ideal implementation of the BB84. Figure (b) exemplifies the acceptance polarization region associated to the BB84 state $\ket{0}$, given by a rectangular solid angle centered in this state with angular widths $2\Delta\phi$ and $2\Delta\theta$.}
	\label{fig:Bloch}
\end{figure}

From Eq.~(\ref{averaging}) to Eq.~(\ref{PDF_2}), it follows that the mixed output state of the transmitter reads
\begin{equation}
\sigma=\frac{1}{2\pi}\int_{-\pi}^{\pi}d\phi\int_{0}^{\pi}d\theta\int_{0}^{I_{\mathrm{max},\theta}}dI{}f_{\boldsymbol{\theta},\boldsymbol{I}}(\theta,I)\sum_{n=0}^{\infty}\frac{e^{-I}{I}^{n}}{n!}\ketbra{n}{n}_{\theta,\phi}
\end{equation}
if no post-selection step is performed. Coming next, we define acceptance regions in the $(\phi,\theta,I)$ space for the post-selection, which are presumed to be accurately identified by the photo-detection system in Fig.~\ref{schematic}. For the standard decoy-state BB84 protocol with three intensity settings, an obvious choice is
\begin{equation}\label{regions}
\Omega_{x,j}=\biggl\{\phi\in\left(x-\Delta\phi,x+\Delta\phi\right),\theta\in\left(\frac{\pi}{2}-\Delta\theta,\frac{\pi}{2}+\Delta\theta\right),I\in{}I_{j}\biggr\},
\end{equation}
where $x\in\{0,\pi,\pi/2,-\pi/2\}$ tags the BB84 polarization states, $\Delta\phi\in(0,\pi/4)$ and $\Delta\theta\in(0,\pi/2)$ define the angular widths of the acceptance regions in the RL Bloch sphere (see Fig.~\ref{fig:Bloch}), and $I_{j}$ stands for the interval of intensities that defines the $j$-th intensity setting, $j\in\{\mathrm{s\ (``signal"),\ d\ (``decoy"),\ v\ (``vacuum")}\}$. As the notation suggests, we shall assume below that the key is extracted from the signal setting, while the decoy and the vacuum settings are only used for parameter estimation.

The resulting post-selected states read $\sigma_{x,j}=\tilde\sigma_{x,j}/\Tr\left[\tilde\sigma_{x,j}\right]$ for
\begin{equation}\label{post-selected}
\tilde\sigma_{x,j}=\frac{1}{2\pi}\int_{x-\Delta\phi}^{x+\Delta\phi}d\phi\int_{\frac{\pi}{2}-\Delta\theta}^{\frac{\pi}{2}+\Delta\theta}d\theta\int_{I_{j}}dI{}f_{\boldsymbol{\theta},\boldsymbol{I}}(\theta,I)\sum_{n=0}^{\infty}\frac{e^{-I}{I}^{n}}{n!}\ketbra{n}{n}_{\theta,\phi}.
\end{equation}
Note that
\begin{equation}\label{p_xj}
\Tr\left[\tilde\sigma_{x,j}\right]=\frac{1}{2\pi}\int_{x-\Delta\phi}^{x+\Delta\phi}d\phi\int_{\frac{\pi}{2}-\Delta\theta}^{\frac{\pi}{2}+\Delta\theta}d\theta\int_{I_{j}}dI{}f_{\boldsymbol{\theta},\boldsymbol{I}}(\theta,I)=\frac{\Delta\phi}{\pi}\int_{\frac{\pi}{2}-\Delta\theta}^{\frac{\pi}{2}+\Delta\theta}d\theta\int_{I_{j}}dI{}f_{\boldsymbol{\theta},\boldsymbol{I}}(\theta,I).
\end{equation}
That is to say, $\Tr\left[\tilde\sigma_{x,j}\right]$ provides the probability that the output state lies in the acceptance region $\Omega_{x,j}$. Notably as well, given the set $\{\Omega_{x,j}\}_{x}$, the $Z$ ($X$) basis acceptance region associated to the $j$-th intensity setting is constructed as $\Omega_{j}^{Z}=\Omega_{0,j}\cup{}\Omega_{\pi,j}$ $(\Omega_{j}^{X}=\Omega_{\frac{\pi}{2},j}\cup{}\Omega_{-\frac{\pi}{2},j})$.\\

In what follows, we shall use the shorthand notation $\left\langle\cdot\right\rangle_{\Omega}$ to denote the triple integral of any input ``$\cdot$" weighted by $f_{\boldsymbol{\phi},\boldsymbol{\theta},\boldsymbol{I}}(\phi,\theta,I)=f_{\boldsymbol{\theta},\boldsymbol{I}}(\theta,I)/2\pi$ in the region $\Omega$ of the $(\phi,\theta,I)$-space. As an example, with this convention Eq.~(\ref{post-selected}) reads
\begin{equation}\label{example}
\tilde\sigma_{x,j}=\left\langle\sum_{n=0}^{\infty}\frac{e^{-I}{I}^{n}}{n!}\ketbra{n}{n}_{\theta,\phi}\right\rangle_{\Omega_{x,j}},
\end{equation}
and $\Tr\left[\tilde\sigma_{x,j}\right]=\left\langle{1}\right\rangle_{\Omega_{x,j}}$. Also, with this notation, it is obvious that $\sigma_{x,j}$ is a convex combination of Fock states. To be precise, normalizing Eq.~(\ref{example}) immediately yields
\begin{equation}\label{example_2}
\sigma_{x,j}=\sum_{n=0}^{\infty}p\left(n|\Omega_{x,j}\right)\sigma_{x,j,n},
\end{equation}
where $p\left(n|\Omega_{x,j}\right)=\left\langle{\frac{e^{-I}{I}^{n}}{n!}}\right\rangle_{\Omega_{x,j}}\Bigl/\bigl\langle{1}\bigr\rangle_{\Omega_{x,j}}$ and
\begin{equation}\label{Fock}
\sigma_{x,j,n}=\frac{\left\langle\frac{e^{-I}{I}^{n}}{n!}\ketbra{n}{n}_{\theta,\phi}\right\rangle_{\Omega_{x,j}}}{\left\langle{\frac{e^{-I}{I}^{n}}{n!}}\right\rangle_{\Omega_{x,j}}}
\end{equation}
is a Fock state with photon number $n$. Notably, the conditional photon-number statistics $p\left(n|\Omega_{x,j}\right)$ are independent of $x$ because of the azimuthal symmetry. Therefore, below we shall denote $p\left(n|\Omega_{x,j}\right)$ simply as $p_{n|j}$ for all $x$.

Similarly, the same symmetry shows that, if we focus on the acceptance region $\Omega_{j}^{Z}=\Omega_{0,j}\cup{}\Omega_{\pi,j}$, the post-selected output state of the transmitter reads
\begin{equation}\label{sigma_z_j}
\sigma_{j}^{Z}=\frac{1}{\left\langle{1}\right\rangle_{\Omega_{j}^{Z}}}{\left\langle\sum_{n=0}^{\infty}\frac{e^{-I}{I}^{n}}{n!}\ketbra{n}{n}_{\theta,\phi}\right\rangle_{\Omega_{j}^{Z}}}=\sum_{n=0}^{\infty}p_{n|j}\sigma_{j,n}^{Z}
\end{equation}
with $\sigma_{j,n}^{Z}=\left(\sigma_{0,j,n}+\sigma_{\pi,j,n}\right)/2$.

Identically, for the $X$ basis we have $\sigma_{j}^{X}=\sum_{n=0}^{\infty}p_{n|j}\sigma_{j,n}^{X}$ with $\sigma_{j,n}^{X}=\left(\sigma_{\frac{\pi}{2},j,n}+\sigma_{-\frac{\pi}{2},j,n}\right)/2$.
\section{Decoy-state analysis}\label{decoy-state}
In this section, we present the decoy-state equations for the $Z$ basis, and the ones for the $X$ basis are discussed at the end.\\

In the first place, it is necessary to introduce some notation. Let $Q^{Z}_{j}$ ($E^{Z}_{j}$) be the probability that a ``click" (an ``error") is recorded conditioned on the event that $\sigma_{j}^{Z}$ is post-selected and Bob performs his measurement in the $Z$ basis. Namely, $Q^{Z}_{j}=p\left(\mathrm{click}|\sigma_{j}^{Z},Z\right)$ and $E^{Z}_{j}=p\left(\mathrm{err}|\sigma_{j}^{Z},Z\right)$. Similarly, let $y^{Z}_{j,n}$ and $e^{Z}_{j,n}$ denote the corresponding $n$-photon yield and $n$-photon error probability, respectively, such that $y^{Z}_{j,n}=p\left(\mathrm{click}|\sigma_{j,n}^{Z},Z\right)$ and $e^{Z}_{j,n}=p\left(\mathrm{err}|\sigma_{j,n}^{Z},Z\right)$.

From the above definitions and Eq.~(\ref{sigma_z_j}), it follows that $Q^{Z}_{j}=\sum_{n=0}^{\infty}p_{n|j}y^{Z}_{j,n}$ and $E^{Z}_{j}=\sum_{n=0}^{\infty}p_{n|j}e^{Z}_{j,n}$ for all $j$. Therefore, truncating the sums to a threshold photon number $n_{\rm cut}$, we find the constraints
\begin{equation}\label{decoy_constraints}
\begin{cases}
Q^{Z}_{j}\geq\displaystyle\sum_{n=0}^{n_{\rm cut}}p_{n|j}y^{Z}_{j,n}, \\
Q^{Z}_{j}\leq\displaystyle\sum_{n=0}^{n_{\rm cut}}p_{n|j}y^{Z}_{j,n}+1-\displaystyle\sum_{n=0}^{n_{\rm cut}}p_{n|j}  
\end{cases}
\hspace{.5cm}\mathrm{and}\hspace{.5cm}
\begin{cases}
E^{Z}_{j}\geq\displaystyle\sum_{n=0}^{n_{\rm cut}}p_{n|j}e^{Z}_{j,n}, \\
E^{Z}_{j}\leq\displaystyle\sum_{n=0}^{n_{\rm cut}}p_{n|j}e^{Z}_{j,n}+1-\displaystyle\sum_{n=0}^{n_{\rm cut}}p_{n|j},
\end{cases}
\end{equation}
for $j\in\{\mathrm{s,d,v}\}$. Setting a threshold photon number $n_{\rm cut}$ as we do allows to define finite linear programs to estimate the single-photon yield and the single-photon error probability of the signal intensity window (\textit{i.e.}, the one devoted to key extraction).

Note that the Fock states $\sigma_{j,n}^{Z}$ and $\sigma_{x,j,n}$ are (generally) different for each setting $j$, and thus distinguishable for Eve. This implies that the yields $y^{Z}_{j,n}$ and the error probabilities $e^{Z}_{j,n}$ might be setting-dependent. As a consequence, it is mandatory to incorporate additional constraints in the decoy-state analysis. For this purpose, the tool that we use is the TD argument~\cite{Nielsen}, presented in Appendix~\ref{TD}. This tool fundamentally limits the maximum bias that Eve may induce between the measurement statistics of two non-orthogonal quantum states, thus naturally providing upper bounds on the differences $\abs{y^{Z}_{j,n}-y^{Z}_{k,n}}$ and $\abs{e^{Z}_{j,n}-e^{Z}_{k,n}}$ for all $j,k\in\{\mathrm{s,d,v}\}$ ($j\neq{}k$) and $n\in\mathbb{N}$. If we denote these bounds respectively as $\Delta_{j,k,n}^{Z}$ and $\tilde\Delta_{j,k,n}^{Z}$, the resulting linear programs that fulfil the decoy-state method read
\begin{eqnarray}\label{lp_1}
&&\min\quad y_{\mathrm{s},1}^{Z}\nonumber\\
&&\textup{s.t.}\hspace{.3cm}Q^{Z}_{j}\geq\displaystyle\sum_{n=0}^{n_{\rm cut}}p_{n|j}y^{Z}_{j,n}\hspace{.2cm}(j\in\{\mathrm{s,d,v}\}),\nonumber\\
&&\hspace{.8cm}Q^{Z}_{j}\leq\displaystyle\sum_{n=0}^{n_{\rm cut}}p_{n|j}y^{Z}_{j,n}+1-\displaystyle\sum_{n=0}^{n_{\rm cut}}p_{n|j} \hspace{.2cm}(j\in\{\mathrm{s,d,v}\}),\nonumber\\
&&\hspace{.8cm}\left\lvert{y^{Z}_{j,n}-y^{Z}_{k,n}}\right\rvert\leq{}\Delta_{j,k,n}^{Z}\hspace{.2cm}(j,k\in\{\mathrm{s,d,v}\},\ j\neq{}k,\ n=0,\ldots{},n_{\rm cut}),\nonumber\\
&&\hspace{.8cm}0\leq{}y^{Z}_{j,n}\leq{}1\hspace{.2cm}(j\in\{\mathrm{s,d,v}\},\ n=0,\ldots,n_{\rm cut}),
\end{eqnarray}
for the signal-setting single-photon yield, and
\begin{eqnarray}\label{lp_2}
&&\max\quad e_{\mathrm{s},1}^{Z}\nonumber\\
&&\textup{s.t.}\hspace{.3cm}E^{Z}_{j}\geq\displaystyle\sum_{n=0}^{n_{\rm cut}}p_{n|j}e^{Z}_{j,n}\hspace{.2cm}(j\in\{\mathrm{s,d,v}\}),\nonumber\\
&&\hspace{.8cm}E^{Z}_{j}\leq\displaystyle\sum_{n=0}^{n_{\rm cut}}p_{n|j}e^{Z}_{j,n}+1-\displaystyle\sum_{n=0}^{n_{\rm cut}}p_{n|j} \hspace{.2cm}(j\in\{\mathrm{s,d,v}\}),\nonumber\\
&&\hspace{.8cm}\left\lvert{e^{Z}_{j,n}-e^{Z}_{k,n}}\right\rvert\leq{}\tilde\Delta_{j,k,n}^{Z}\hspace{.2cm}(j,k\in\{\mathrm{s,d,v}\},\ j\neq{}k,\ n=0,\ldots{},n_{\rm cut}),\nonumber\\
&&\hspace{.8cm}0\leq{}e^{Z}_{j,n}\leq{}1\hspace{.2cm}(j\in\{\mathrm{s,d,v}\},\ n=0,\ldots,n_{\rm cut}),
\end{eqnarray}
for the signal-setting single-photon error probability.\\

Importantly, replacing $Z$ by $X$ everywhere above, the relevant quantities ---$Q^{X}_{j}$, $E^{X}_{j}$, $y^{X}_{j,n}$, $e^{X}_{j,n}$, $\Delta_{j,k,n}^{X}$ and $\tilde\Delta_{j,k,n}^{X}$--- and linear programs for the $X$ basis follow. Coming next, we compute the specific values of $\Delta_{j,k,n}^{Z}$, $\Delta_{j,k,n}^{X}$, $\tilde\Delta_{j,k,n}^{Z}$ and $\tilde\Delta_{j,k,n}^{X}$  using the TD argument.
\subsection{Calculation of $\Delta_{j,k,n}^{Z}$ and $\Delta_{j,k,n}^{X}$}
For simplicity of the notation, the derivation below assumes collective attacks. Nevertheless, explicit calculation easily shows that the resulting bounds hold against fully general attacks too. This feature follows from the tensor product structure of the global state of all protocol rounds.\\

Let $\hat{U}_{\rm BE}$ denote Eve's unitary operation in any given round, acting on the system B transmitted through the channel and a probe system E under Eve's control, initialized in a certain state $\ket{\varphi}_{\rm E}$. Also, for the purpose of evaluating the yields, Bob's possible measurement outcomes are either ``click" or ``no click". Therefore, his $Z$ basis measurement is described by a positive-operator-valued measure (POVM) with elements $\{\hat{M}^{\mathrm{click}}_{\rm B},\hat{M}^{\mathrm{no}\hspace{.05cm}\mathrm{click}}_{\rm B}\}$, where $\hat{M}^{\mathrm{no}\hspace{.05cm}\mathrm{click}}_{\rm B}=\mathds{1}_{\rm B}-\hat{M}^{\mathrm{click}}_{\rm B}$. Note that we are assuming the basis-independent detection efficiency condition here, such that no basis dependence is included in the POVM elements.

Now, recalling that $y^{Z}_{j,n}=p\left(\mathrm{click}|\sigma_{j,n}^{Z},Z\right)$, it follows that
\begin{equation}\label{yield}
y^{Z}_{j,n}=\Tr\left[\hat{U}^{\dagger}_{\rm BE}\hat{M}^{\mathrm{click}}_{\rm B}\hat{U}_{\rm BE}\Bigl(\sigma_{j,n}^{Z}\otimes\ketbra{\varphi}{\varphi}_{\rm E}\Bigr)\right]
\end{equation}
for all $j\in\{\mathrm{s,d,v}\}$ and $n\in\mathbb{N}$. Note that, aiming to keep the notation introduced in Sec.~\ref{Characterization}, the subscript B of Bob's system is not made explicit in the state between brackets. From Eq.~(\ref{yield}), direct application of the TD argument (see Appendix~\ref{TD}) yields
\begin{equation}
\left\lvert{y^{Z}_{j,n}-y^{Z}_{k,n}}\right\rvert\leq{}D\left(\sigma_{j,n}^{Z},\sigma_{k,n}^{Z}\right)
\end{equation}
for all $j,k\in\{\mathrm{s,d,v}\}$ and $n\in\mathbb{N}$, where
\begin{equation}
D\left(\sigma_{j,n}^{Z},\sigma_{k,n}^{Z}\right)=\frac{1}{2}\Tr\left[\sqrt{\left(\sigma_{j,n}^{Z}-\sigma_{k,n}^{Z}\right)^{2}}\right]=\Delta_{j,k,n}^{Z}
\end{equation}
is the TD between $\sigma_{j,n}^{Z}$ and $\sigma_{k,n}^{Z}$.

A similar procedure leads to $\Delta_{j,k,n}^{X}=D\bigl(\sigma_{j,n}^{X},\sigma_{k,n}^{X}\bigr)$, and in virtue of the azimuthal symmetry it follows that $\Delta_{j,k,n}^{X}=\Delta_{j,k,n}^{Z}$ for all possible inputs. To finish with, we remark that evaluating the TD values $\Delta_{j,k,n}^{Z}$ requires to provide a matrix representation of the input density matrices. In this regard, a natural representation is given in Appendix~\ref{representation}.
\subsection{Calculation of $\tilde\Delta_{j,k,n}^{Z}$ and $\tilde\Delta_{j,k,n}^{X}$}
For the purpose of evaluating the $Z$ basis error probabilities, finer-grained measurement operators are required, in a one-to-one correspondence with Bob's possible outcomes ``$0$", ``$\pi$" and ``no click" (as usual, double clicks are randomly assigned to a detection event, \textit{i.e.}, ``$0$" or ``$\pi$" in this case). Therefore, error-wise, Bob's measurement is described by a POVM with elements $\{\hat{M}^{0}_{\rm B},\hat{M}^{\pi}_{\rm B},\hat{M}^{\mathrm{no}\hspace{.05cm}\mathrm{click}}_{\rm B}\}$, where $\hat{M}^{0}_{\rm B}+\hat{M}^{\pi}_{\rm B}=\hat{M}^{\mathrm{click}}_{\rm B}$.

Recalling that $e^{Z}_{j,n}=p\left(\mathrm{err}|\sigma_{j,n}^{Z},Z\right)$ and that $\sigma_{j,n}^{Z}=\left(\sigma_{0,j,n}+\sigma_{\pi,j,n}\right)/2$, it follows that
\begin{eqnarray}\label{error}
&&e^{Z}_{j,n}=\frac{1}{2}\bigl[p\left(\mathrm{err}|\sigma_{0,j,n},Z\right)+p\left(\mathrm{err}|\sigma_{\pi,j,n},Z\right)\bigr]=\nonumber \\
&&=\frac{1}{2}\biggl\{\Tr\left[\hat{U}^{\dagger}_{\rm BE}\hat{M}^{\pi}_{\rm B}\hat{U}_{\rm BE}\Bigl(\sigma_{0,j,n}\otimes\ketbra{\varphi}{\varphi}_{\rm E}\Bigr)\right]+\Tr\left[\hat{U}^{\dagger}_{\rm BE}\hat{M}^{0}_{\rm B}\hat{U}_{\rm BE}\Bigl(\sigma_{\pi,j,n}\otimes\ketbra{\varphi}{\varphi}_{\rm E}\Bigr)\right]\biggr\}
\end{eqnarray}
for all $j\in\{\mathrm{s,d,v}\}$ and $n\in\mathbb{N}$. That is to say, upon post-selection of $\sigma_{0,j,n}$ ($\sigma_{\pi,j,n}$), an error occurs if Bob records the outcome ``$\pi$" (``$0$"). Hence, defining $e_{0,j,n}=p\left(\mathrm{err}|\sigma_{0,j,n},Z\right)$ and $e_{\pi,j,n}=p\left(\mathrm{err}|\sigma_{\pi,j,n},Z\right)$, the TD argument provides the constraints
\begin{equation}\label{bias_errors}
\abs{e_{0,j,n}-e_{0,k,n}}\leq{}D\left(\sigma_{0,j,n},\sigma_{0,k,n}\right)\hspace{.5cm}\mathrm{and}\hspace{.5cm}\abs{e_{\pi,j,n}-e_{\pi,k,n}}\leq{}D\left(\sigma_{\pi,j,n},\sigma_{\pi,k,n}\right)
\end{equation}
for all $j,k\in\{\mathrm{s,d,v}\}$ and $n\in\mathbb{N}$. From Eq.~(\ref{bias_errors}) and the triangle inequality, the desired bound on the bias $\abs{e^{Z}_{j,n}-e^{Z}_{k,n}}$ follows. Namely,
\begin{equation}
\left\lvert{e^{Z}_{j,n}-e^{Z}_{k,n}}\right\rvert\leq{}\frac{1}{2}\bigl[D\left(\sigma_{0,j,n},\sigma_{0,k,n}\right)+D\left(\sigma_{\pi,j,n},\sigma_{\pi,k,n}\right)\bigr]
\end{equation}
for all $j,k\in\{\mathrm{s,d,v}\}$ and $n\in\mathbb{N}$. In conclusion, $\tilde\Delta_{j,k,n}^{Z}=\bigl[D\left(\sigma_{0,j,n},\sigma_{0,k,n}\right)+D\left(\sigma_{\pi,j,n},\sigma_{\pi,k,n}\right)\bigr]\bigl/2$. What is more, the azimuthal symmetry assures that $D\left(\sigma_{0,j,n},\sigma_{0,k,n}\right)=D\left(\sigma_{\pi,j,n},\sigma_{\pi,k,n}\right)$, and thus
\begin{equation}
\tilde\Delta_{j,k,n}^{Z}=D\left(\sigma_{0,j,n},\sigma_{0,k,n}\right).
\end{equation}

Again, proceeding identically and invoking the symmetry, for the $X$ basis one finds $\tilde\Delta_{j,k,n}^{X}=D\left(\sigma_{\frac{\pi}{2},j,n},\sigma_{\frac{\pi}{2},k,n}\right)=D\left(\sigma_{0,j,n},\sigma_{0,k,n}\right)=\tilde\Delta_{j,k,n}^{Z}$ for all possible inputs. As before, the reader is referred to Appendix~\ref{representation} for a matrix representation of the $\sigma_{x,j,n}$, necessary for the calculation of the TD values $\tilde\Delta_{j,k,n}^{Z}$.
\section{Entanglement-based protocol and single-photon phase-error rate}\label{PHER}
As in the previous section, we assume the restricted scenario of collective attacks, and recall that this suffices to establish a valid asymptotic key rate analysis against coherent attacks in virtue of the de Finetti theorem~\cite{Renner1} or the post-selection technique~\cite{Renner2}.

Throughout this section, we shall only refer to the single-photon component $\sigma_{x,\mathrm{s},1}$ of the output states $\sigma_{x,\mathrm{s}}$ (see Eq.~(\ref{Fock})), where we recall that we set $j= \rm s$ because we assume that the key is extracted from the signal intensity window, $I_{\mathrm{s}}$. For convenience, we define $\ket{\rm R}=a^{\dagger}_{\rm R}\ket{\rm vac}$ and $\ket{\rm L}=a^{\dagger}_{\rm L}\ket{\rm vac}$ (poles of the RL Bloch sphere).

The formulation of the virtual entanglement-based (EB) protocol proceeds in two steps. In a first step, we consider a purification of $\sigma_{x,\mathrm{s},1}$ via a shield qubit system A inaccessible to all Alice, Bob and Eve. In particular, this requires diagonalizing $\sigma_{x,\mathrm{s},1}$ first. In a second step, we consider an additional purification via an ancillary system $\rm A'$ held by Alice, which determines the states she prepares for Bob via projective measurements as usual. Bob's system shall be denoted by the subscript B.

For the purpose of diagonalizing $\sigma_{x,\mathrm{s},1}=\langle{e^{-I}{I}\ketbra{1}{1}_{\theta,\phi}}\rangle_{\Omega_{x,\mathrm{s}}}/\langle{e^{-I}{I}}\rangle_{\Omega_{x,\mathrm{s}}}$, we use again the matrix representation provided in Appendix~\ref{representation}. In particular, from Eq.~(\ref{matlab_input}), straightforward algebra leads to
\begin{equation}\label{computational_basis}
\sigma_{x,\mathrm{s},1}=\frac{\mathds{1}_{\rm B}}{2}+\Delta_{\mathrm{s}}\left(e^{ix}\ketbra{\rm L}{\rm R}_{\rm B}+e^{-ix}\ketbra{\rm R}{\rm L}_{\rm B}\right)
\end{equation}
in the orthonormal basis $\{\ket{\rm R},\ket{\rm L}\}$, where $\mathds{1}_{\rm B}$ denotes the identity operator and
\begin{equation}
\Delta_{\mathrm{s}}=\frac{\sin(\Delta\phi)}{2\Delta\phi}\times\frac{\displaystyle{\int_{\frac{\pi}{2}-\Delta\theta}^{\frac{\pi}{2}+\Delta\theta}d\theta\sin\theta\int_{I_{\mathrm{s}}}dI{}f_{\boldsymbol{\theta},\boldsymbol{I}}(\theta,I)e^{-I}{I}}}{\displaystyle{\int_{\frac{\pi}{2}-\Delta\theta}^{\frac{\pi}{2}+\Delta\theta}d\theta\int_{I_{\mathrm{s}}}dI{}f_{\boldsymbol{\theta},\boldsymbol{I}}(\theta,I)e^{-I}{I}}}.
\end{equation}
Naturally, $\sigma_{x,\mathrm{s},1}$ is diagonal in the $\{\ket{x}_{\rm B},\ket{x+\pi}_{\rm B}\}$ basis, where $\ket{x}_{\rm B}=\frac{1}{\sqrt{2}}\left(\ket{\rm R}_{\rm B}+e^{ix}\ket{\rm L}_{\rm B}\right)$. Note that $\ket{x}_{\rm B}$ is a pure state in the RL Bloch sphere with $\theta=\pi/2$ and $\phi=x$. In particular, explicit diagonalization yields
\begin{equation}
\sigma_{x,\mathrm{s},1}=\left(\frac{1}{2}+\Delta_{\mathrm{s}}\right)\ketbra{x}{x}_{\rm B}+\left(\frac{1}{2}-\Delta_{\mathrm{s}}\right)\ketbra{x+\pi}{x+\pi}_{\rm B},
\end{equation}
such that after attaching the shield system A ---with, say, orthonormal basis $\{\ket{0}_{\rm A},\ket{\pi}_{\rm A}\}$---, the purified state reads
\begin{equation}\label{shield}
\ket{\Psi_{x,\mathrm{s},1}}_{\rm AB}=\sum_{\delta\in\{0,\pi\}}\left(\frac{1}{2}+e^{i\delta}\Delta_{\mathrm{s}}\right)^{1/2}\ket{\delta}_{\rm A}\ket{x+\delta}_{\rm B}.
\end{equation}

As stated above, in the virtual EB approach Alice holds an ancillary polarization qubit $\rm A'$ maximally entangled to Bob's purified qubit AB. That is to say, whenever $\Omega^{Z}_{\mathrm{s}}$ is post-selected and a single-photon is emitted, the equivalent three-partite state prepared by Alice in the EB protocol reads
\begin{equation}\label{computational}
\ket{\Psi_{\mathrm{s},1}}_{\rm A'AB}=\frac{1}{\sqrt{2}}\Bigl(\ket{0}_{\mathrm{A'}}\ket{\Psi_{0,\mathrm{s},1}}_{\rm AB}+\ket{\pi}_{\mathrm{A'}}\ket{\Psi_{\pi,\mathrm{s},1}}_{\rm AB}\Bigr),
\end{equation}
where again we are using the RL Bloch sphere notation $\ket{y}_{\rm A'}=\frac{1}{\sqrt{2}}\left(\ket{\rm R}_{\rm A'}+e^{iy}\ket{\rm L}_{\rm A'}\right)$. Importantly, it is shown in Appendix~\ref{EB} that Eq.~(\ref{computational}) can be rewritten as
\begin{equation}\label{test}
\ket{\Psi_{\mathrm{s},1}}_{\rm A'AB}=\frac{1}{\sqrt{2}}\left(\ket{\frac{\pi}{2}}_{\mathrm{A'}}\ket{\Psi_{-\frac{\pi}{2},\mathrm{s},1}}_{\rm AB}+\ket{-\frac{\pi}{2}}_{\mathrm{A'}}\ket{\Psi_{\frac{\pi}{2},\mathrm{s},1}}_{\rm AB}\right).
\end{equation}
Therefore, the $Z$ basis phase-error probability $\phi^{Z}_{\rm s}$ is defined as the bit-error probability between the outcomes $\textbf{X}_{\rm A'}$ and $\textbf{X}_{\rm B}$, reached by measuring the polarization qubits $\rm A'$ and $\rm B$ in the test bases~\cite{uncertainty} (\textit{i.e.}, $\{\ketbra{\pi/2}{\pi/2}_{\rm A'},\ketbra{-\pi/2}{-\pi/2}_{\rm A'}\}$ and $\{\ketbra{\pi/2}{\pi/2}_{\rm B},\ketbra{-\pi/2}{-\pi/2}_{\rm B}\}$, respectively). In the prepare-and-measure picture, this matches the bit-error probability that arises when $\Omega^{X}_{\mathrm{s}}$ is post-selected, a single-photon is emitted (these two features are equivalent to asserting that $\sigma_{\mathrm{s},1}^{X}$ is post-selected), and Bob selects the $X$ basis. Namely,
\begin{equation}\label{phase_error}
\phi^{Z}_{\rm s}=\frac{p\left(\mathrm{err}|\sigma_{\mathrm{s},1}^{X},X\right)}{p\left(\mathrm{click}|\sigma_{\mathrm{s},1}^{X},X\right)}=\frac{e_{\mathrm{s},1}^{X}}{y_{\mathrm{s},1}^{X}}.
\end{equation}
As usual, one can define the $X$ basis single-photon phase-error probability $\phi^{X}_{\rm s}$ in an entirely identical fashion, and it can be computed as $\phi^{X}_{\rm s}={e_{\mathrm{s},1}^{Z}}/{y_{\mathrm{s},1}^{Z}}$ following the same argument presented here.
%
%
%
\section{Performance}\label{Performance}
The secret key rate formula of our passive QKD scheme is determined by the fact that we consider a decoy-state BB84 protocol, with the minor difference that one must deal with the continuous post-selection regions introduced by the PT.

Naturally, we assume that both bases are used for key extraction, because they are equally likely to be post-selected. In short, the secret key rate reads $K=K_{Z}+K_{X}$ with
\begin{equation}
K_{M}=q_{M}\times\left\{\left\langle{e^{-I}{I}}\right\rangle_{\Omega_{\rm s}^{M}}y^{M}_{\mathrm{s},1}\Bigl[1-h\left(\phi^{M}_{\rm s}\right)\Bigr]-f_{\rm EC}\left\langle{1}\right\rangle_{\Omega_{\rm s}^{M}}Q^{M}_{\rm s}h\left(\frac{E^{M}_{\rm s}}{Q^{M}_{\rm s}}\right)\right\},
\end{equation}
where $M\in\{Z,X\}$, $q_{M}$ stands for Bob's probability to select basis $M$ ($q_{M}=1/2$ being optimal for symmetry reasons), $h(\cdot)$ stands for Shannon's binary entropy function, and $f_{\rm EC}$ denotes the error correction efficiency.

\begin{figure}[!htbp]
	\centering 
	\includegraphics[width=11.4cm,height=7.7cm]{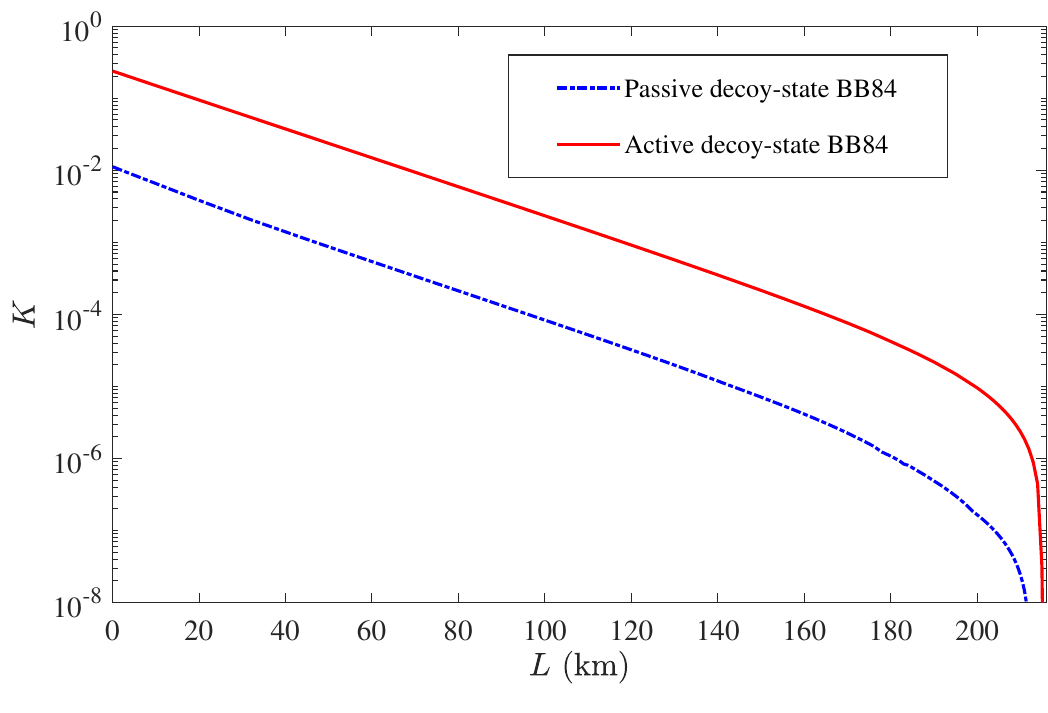}\\
	\caption{Rate-distance performance of our fully passive decoy-state BB84 proposal. For the sake of comparison, we include the secret key rate attained in the active setting as well (solid red line). The experimental parameters are set as specified in the main text.}
	\label{fig:performance}
\end{figure}

In what follows, we evaluate the rate-distance performance of the PT illustrated in Fig.~\ref{schematic}. For this purpose, in the absence of experimental data, we consider a natural channel and detector model presented in Appendix~\ref{channel_model}. The model is specified by the channel transmittance, $\eta_{\rm ch}=10^{-\alpha_{\rm att}{}L/10}$ ---where $\alpha_{\rm att}$ denotes the attenuation coefficient of the channel and $L$ stands for its transmission length---, the detector efficiency of Bob's detectors, $\eta_{\rm Bob}$, and their dark count rate, $p_{\rm d}$. For illustration purposes, we set these parameters to typical values of $\alpha_{\rm att}=0.2$ dB/km (telecom wavelength attenuation), $\eta_{\rm Bob}=65\%$ and $p_{\rm d}=10^{-6}$. As for the input settings of the PT, we assume that no intensity value in the accessible range $(0,4\nu{}t)$ is withdrawn, such that the intervals $I_{\rm v}$, $I_{\rm d}$ and $I_{\rm s}$ are strictly consecutive and exhaustive in this range. Also, we fix the width of $I_{\rm v}/4\nu{}t$ and $I_{\rm d}/4\nu{}t$ to a reasonable small value of $5\times{}10^{-3}$ for the numerics. Hence, $I_{\rm v}/4\nu{}t=(0,5)\times{}10^{-3}$ and $I_{\rm d}/4\nu{}t=(5,10)\times{}10^{-3}$. The product $\nu{}t$ and the angular widths $\Delta\theta$ and $\Delta\phi$ of the post-selection regions are numerically optimized to maximize the secret key rate for each value of $L$, and the optimization reveals a roughly constant optimal value $\nu{}t\approx{}0.25$. Finally, we set the threshold photon number for the decoy-state linear programs to $n_{\rm cut}=3$. Importantly, the loss of generality of the above numerical specifications is very small, as we check that the delivered secret key rate is remarkably close to the perfect decoy-state parameter estimation limit (which we compute under full optimization as well).
The results are shown in Fig.~\ref{fig:performance}, where we further include the secret key rate reached in the active setting for comparison purposes. 

The figure shows that the security and simplicity upgrades of the PT come at the price of lowering the secret key rate by a factor $\sim{}1/20$. As mentioned in Sec.~\ref{Introduction}, the origin of this discrepancy is twofold. In the first place, the post-selection of acceptance regions requires additional sifting compared to the active setting. In the second place, the post-selected $\sigma_{x,j,n}$ are in a mixed polarization state. This represents an inherent source of noise not present in the active case, where one typically considers pure states with fixed polarizations.
\section{Discussion}\label{Discussion}
Active QKD systems often rely on the use of externally driven modulators, whose imperfections lead to security loopholes and side-channels (\textit{e.g.} THAs), and whose frequency of operation typically limits the repetition rate of the system. Therefore, replacing all externally driven elements by a passive mechanism could be a very appealing feature for QKD: it provides immunity to modulator side channels and, in principle, it enables the operation of QKD systems at higher repetition rates. On top of it, passive architectures could reduce the complexity (and thereby the cost) of QKD infrastructures. Needless to say, this would entail an advantage in many practical situations, for instance, when it comes to deploying QKD on a satellite.

Despite the intense research in the field, working out a fully passive linear-optics setup for the celebrated decoy-state BB84 protocol remained an open problem for more than a decade. Nonetheless, the present work, together with that in~\cite{Mike}, elucidate this possibility for the first time, and explicitly construct different fully passive protocols with very diverse approaches to polarization encoding and decoy-state parameter estimation. This clearly illustrates the versatility of the considered prototype. On top of it, we provide a detailed security analysis for the mixed single-photon states generated by the transmitter, tightening a loose end present in~\cite{Mike}.

For implementation purposes, we remark that the usage of four independent lasers in the PT is not a necessity by any means, but rather an instrument we use for theoretical convenience. Identically as in~\cite{Mike} ---where a detailed discussion is included---, the same physical states can be generated by using a single laser and suitable Mach-Zehnder interferometry. This configuration benefits from an enhanced simplicity, because high-visibility interference demands photon indistinguishability, which is typically hard to enforce when using independent lasers.

Furthermore, despite the resilience to modulator side-channels or THAs, certain other vulnerabilities may affect a PT. For instance, Eve could try to perform a laser-seeding attack to modify the phase/intensity of the laser pulses generated in the PT~\cite{Sun1,Sun2,Huang2}. Attacks of this kind, known to threaten actively modulated systems too, could invalidate the estimation of the secret key length in the passive scenario. What is more, in the case of a PT, Eve could launch a laser-seeding attack to alter the measurement outcome of its photo-detection system. To counter these problems, optical isolation must be incorporated at the output port of the transmitter, as it is done to protect active systems against attacks that inject light from the channel (like \textit{e.g.} THAs). Notwithstanding, current security proofs against THAs in the active setting typically require the intensity of the back-reflected light to be minuscule for the information leakage to be irrelevant (as an example, the analyses in~\cite{Tamaki,Weilong,Navs} demand such intensity to lie below $10^{-7}$ or $10^{-8}$ photons/pulse). In contrast to this, much less isolation is expected to be required to deal with laser-seeding attacks in the PT. This is so because ``classical" (high intensity) light pulses are generated in the PT and arrive at Alice's measurement unit, say, containing $10^{6}$-$10^{8}$ photons/signal. Therefore, as long as the intensity of the injected light is attenuated well below this level, its effect on the generated pulses or on the reading of the detection scheme will probably be negligible. Moreover, any slight modification of the actual measurement outcome due to Eve's action could be readily incorporated in the security proof. Similarly, laser-damage attacks~\cite{Huang} ---which, again, have been proposed against active systems--- may as well jeopardize the security of the PT. An attack of this kind might try to reduce the optical isolation of the transmitter, or to manipulate the behaviour of its internal components. As in the active setting, protection against laser-damage may be achieved with optical isolators, circulators, filters or even an optical fuse~\cite{Zhang,Posonova}. In any case, a detailed analysis of these and other potential attacks where Eve actively meddles with the hardware lies beyond the scope of this work.

Leaving active tampering aside, potential information leakage via back-flash emission from the detection system might be another weakness of a PT that deserves further experimental investigation. If needed, this could probably be circumvented by simply using an anti-reflective coating. Also, we remark that Alice's intensity and polarization measurements are presumed to be noiseless in our analysis. However, to provide protection against a noisy measurement, the noise must be characterized to a certain extent because it affects the post-selected light pulses. Once characterized, it could be incorporated to the security analysis using similar techniques as it is done in the active setting to deal with \textit{e.g.} state preparation flaws or intensity fluctuations.

After these various pending tasks are properly addressed, passive schemes could play a crucial role in the development of practical and affordable QKD solutions, in view of the increasing concerns related to the implementation security of QKD. This being the case, the present work is a valuable input to the topic.
\section{Acknowledgements}\label{Acknowledgements}
All authors gratefully acknowledge Chengqiu Hu for useful discussions. VZ and MC acknowledge support from the Galician Regional Government (consolidation of Research Units: AtlantTIC), the Spanish Ministry of Economy and Competitiveness (MINECO), the Fondo Europeo de Desarrollo Regional (FEDER) through Grant No. PID2020-118178RB-C21, Cisco Systems Inc., and MICINN ---with funding from the European Union NextGenerationEU (PRTR-C17.I1)--- and the Galician Regional Government ---with own funding--- through the “Planes Complementarios de I+D+I con las Comunidades Autonomas” in Quantum Communication. WW is supported by the University of Hong Kong Seed Fund for Basic Research for New Staff and the Hong Kong RGC General Research Fund.

\newpage
\appendix
\section{Calculation of the pure output state of the transmitter}\label{quantum_optics}
Here, we derive Eq.~(\ref{not_randomised}) and Eq.~(\ref{variables}) using elementary quantum optics. Let us consider, say, the top arm of the PT in Fig.~\ref{schematic}. Given $\nu\in\mathbb{R}^{+}$, $\alpha\in[0,2\pi)$ and $\beta\in[0,2\pi)$, the input state of the 50:50 BS reads
\begin{equation}\label{input}
\ket{\varphi_{\rm in}}_{ab}=\ket{\sqrt{\nu}e^{i\alpha}}_{a,\rm R}\ket{\sqrt{\nu}e^{i\beta}}_{b,\rm R}=\exp\left\{\sqrt{\nu}e^{i\alpha}{}a_{\rm R}^{\dagger}-\sqrt{\nu}e^{-i\alpha}{}a_{\rm R}\right\}\times\exp\left\{\sqrt{\nu}e^{i\beta}{}b_{\rm R}^{\dagger}-\sqrt{\nu}e^{-i\beta}{}b_{\rm R}\right\}\ket{\rm vac},
\end{equation}
and the unitary transformation of the BS is given by $\left\{a_{\rm R}^{\dagger}\rightarrow{}\frac{1}{\sqrt{2}}\left(c_{\rm R}^{\dagger}+d_{\rm R}^{\dagger}\right),\ b_{\rm R}^{\dagger}\rightarrow{}\frac{1}{\sqrt{2}}\left(d_{\rm R}^{\dagger}-c_{\rm R}^{\dagger}\right)\right\}$. This leads to
\begin{equation}\label{output}
\ket{\varphi_{\rm out}}_{cd}=\exp\left\{\sqrt{\nu}\left(\frac{e^{i\alpha}-e^{i\beta}}{\sqrt{2}}{}c_{\rm R}^{\dagger}-\frac{e^{-i\alpha}-e^{-i\beta}}{\sqrt{2}}{}c_{\rm R}\right)\right\}\times\exp\left\{\sqrt{\nu}\left(\frac{e^{i\alpha}+e^{i\beta}}{\sqrt{2}}{}d_{\rm R}^{\dagger}-\frac{e^{-i\alpha}+e^{-i\beta}}{\sqrt{2}}{}d_{\rm R}\right)\right\}\ket{\rm vac}
\end{equation}
in the output modes $c$ and $d$, after regrouping terms. Since mode $d$ is not used, from now on we focus on mode $c$, and refer to the corresponding state in Eq.~(\ref{output}) as $\ket{\varphi_{\rm out}}_{c}$. Factoring $\left(e^{i\alpha}-e^{i\beta}\right)/\sqrt{2}$ into modulus and phase yields $\left\lvert{\left(e^{i\alpha}-e^{i\beta}\right)/\sqrt{2}}\right\rvert=\sqrt{1-\cos(\beta-\alpha)}$ and
\begin{equation}\label{phase}
\begin{split}
&\mathrm{Arg}\left(\frac{e^{i\alpha}-e^{i\beta}}{\sqrt{2}}\right)=\frac{e^{i\alpha}-e^{i\beta}}{\sqrt{2\bigl[1-\cos(\beta-\alpha)\bigr]}}=\frac{\cos\alpha-\cos\beta}{2\left\lvert{\sin\left(\frac{\beta-\alpha}{2}\right)}\right\rvert}+i\frac{\sin\alpha-\sin\beta}{2\left\lvert{\sin\left(\frac{\beta-\alpha}{2}\right)}\right\rvert}=\\
&\left[\sin\left(\frac{\alpha+\beta}{2}\right)-i\cos\left(\frac{\alpha+\beta}{2}\right)\right]\mathrm{sgn}(\beta-\alpha),
\end{split}
\end{equation}
where the identity $\sqrt{2\bigl[1-\cos(\beta-\alpha)\bigr]}=2\left\lvert{\sin\left(\frac{\beta-\alpha}{2}\right)}\right\rvert$ is used in the second equality, the identities $\cos\alpha-\cos\beta=2\sin\left(\frac{\alpha+\beta}{2}\right)\sin\left(\frac{\beta-\alpha}{2}\right)$ and $\sin\alpha-\sin\beta=-2\cos\left(\frac{\alpha+\beta}{2}\right)\sin\left(\frac{\beta-\alpha}{2}\right)$ are used in the third equality, and $\mathrm{sgn}(x)$ denotes the sign function. Now, separately addressing the cases $\beta\geq{}\alpha$ and $\beta<\alpha$ in Eq~(\ref{phase}) one obtains
\begin{equation}\label{phase_2}
\mathrm{Arg}\left(\frac{e^{i\alpha}-e^{i\beta}}{\sqrt{2}}\right)=e^{\displaystyle{i\left[\frac{\alpha+\beta}{2}-\mathrm{sgn}(\beta-\alpha)\frac{\pi}{2}\right]}}.
\end{equation}
Putting it all together,
\begin{eqnarray}\label{top_arm}
&&\ket{\varphi_{\rm out}}_{c}=\exp\left\{\sqrt{\nu\bigl[1-\cos(\beta-\alpha)\bigr]}\left[e^{i\left(\frac{\alpha+\beta}{2}-\mathrm{sgn}(\beta-\alpha)\frac{\pi}{2}\right)}c_{\rm R}^{\dagger}-e^{-i\left(\frac{\alpha+\beta}{2}-\mathrm{sgn}(\beta-\alpha)\frac{\pi}{2}\right)}c_{\rm R}\right]\right\}\ket{\rm vac}=\nonumber \\
&&\ket{\sqrt{\nu\bigl[1-\cos(\beta-\alpha)\bigr]}e^{i\left(\frac{\alpha+\beta}{2}-\mathrm{sgn}(\beta-\alpha)\frac{\pi}{2}\right)}}_{c,\rm R}.
\end{eqnarray}
Similarly, the output state of the bottom arm in mode $r$ reads $\ket{\psi_{\rm out}}_{r}=\ket{\sqrt{\nu\bigl[1-\cos(\delta-\gamma)\bigr]}e^{i\left(\frac{\gamma+\delta}{2}-\mathrm{sgn}(\delta-\gamma)\frac{\pi}{2}\right)}}_{r,\rm L}$.

Next, $\ket{\varphi_{\rm out}}_{c}$ and $\ket{\psi_{\rm out}}_{r}$ interfere at a PBS that maps $c_{\rm R}^{\dagger}$ to $v_{\rm R}^{\dagger}$ and $r_{\rm L}^{\dagger}$ to $v_{\rm L}^{\dagger}$. Explicit calculation shows that the output state of the PBS at mode $v$ reads
\begin{equation}\label{PBS}
\ket{\Upsilon}_{v}=\exp\left\{\sqrt{\nu_{\alpha\beta}+\nu_{\gamma\delta}}\ e^{i\tau_{\alpha\beta}}\left[\sqrt{\frac{\nu_{\alpha\beta}}{\nu_{\alpha\beta}+\nu_{\gamma\delta}}}v_{\rm R}^{\dagger}+\sqrt{\frac{\nu_{\gamma\delta}}{\nu_{\alpha\beta}+\nu_{\gamma\delta}}}e^{i\left(\tau_{\gamma\delta}-\tau_{\alpha\beta}\right)}v_{\rm L}^{\dagger}\right]-\widehat{\mathrm{h.c.}}\right\}\ket{\rm vac},
\end{equation}
where we have defined $\nu_{\rho\sigma}=\nu\bigl[1-\cos(\sigma-\rho)\bigr]$ and $\tau_{\rho\sigma}=(\rho+\sigma)/2-\mathrm{sgn}(\sigma-\rho)\pi/2$ for $\rho\in[0,2\pi)$ and $\sigma\in[0,2\pi)$. Also, we introduce the shorthand notation ``$\widehat{\mathrm{h.c.}}$" to denote the hermitian conjugate of the first term between keys.

Eq.~(\ref{PBS}) triggers the definition of $I'=\nu_{\alpha\beta}+\nu_{\gamma\delta}$, $\psi=\tau_{\alpha\beta}$ and $\theta\in[0,\pi]$, $\phi\in[0,2\pi)$ such that $\theta/2=\arctan{\sqrt{\nu_{\gamma\delta}/\nu_{\alpha\beta}}}\hspace{.2cm}\mathrm{and}\hspace{.2cm}\phi=\tau_{\gamma\delta}-\tau_{\alpha\beta}$. In terms of  $I'$, $\psi$, $\theta$ and $\phi$, $\ket{\Upsilon}_{v}$ reads
\begin{equation}
\ket{\Upsilon}_{v}=\exp\left\{\sqrt{I'}e^{i\psi}\left[\cos\left(\frac{\theta}{2}\right)v_{\rm R}^{\dagger}+\sin\left(\frac{\theta}{2}\right)e^{i\phi}v_{\rm L}^{\dagger}\right]-\widehat{\mathrm{h.c.}}\right\}\ket{\rm vac}=\exp\left\{\sqrt{I'}e^{i\psi}v_{\theta,\phi}^{\dagger}-\widehat{\mathrm{h.c.}}\right\}\ket{\rm vac}=\ket{\sqrt{I'}\ e^{i\psi}}_{v,\theta,\phi},
\end{equation}
where we recall that the notation $v_{\theta,\phi}^{\dagger}$ is presented in Eq.~(\ref{creation_operator}). Lastly, $\ket{\Upsilon}_{v}$ enters a BS with transmittance $t$, which maps $v_{\theta,\phi}^{\dagger}$ to $(\sqrt{t}\ w_{\theta,\phi}^{\dagger}+\sqrt{1-t}\ y_{\theta,\phi}^{\dagger})$. This trivially leads to the final output state $\ket{\Psi}_{wy}=\ket{\sqrt{tI'}\ e^{i\psi}}_{w,\theta,\phi}\ket{\sqrt{(1-t)I'}\ e^{i\psi}}_{y,\theta,\phi}$ in modes $w$ and $y$. In particular, setting $I=tI'$, the state that is sent to the channel reads
\begin{equation}\label{finally}
\ket{\Psi}_{w}=\ket{\sqrt{I}\ e^{i\psi}}_{w,\theta,\phi}.
\end{equation}
Making the dependence on $\nu$, $\alpha$, $\beta$, $\gamma$, $\delta$ and $t$ explicit, we see that
\begin{eqnarray}\label{parameters}
&&I=\nu{}t\Bigl[2-\cos(\beta-\alpha)-\cos(\delta-\gamma)\Bigr]=2\nu{}t\left[\sin^{2}\left(\frac{\beta-\alpha}{2}\right)+\sin^{2}\left(\frac{\delta-\gamma}{2}\right)\right],\hspace{1.2cm}\psi=\frac{\alpha+\beta}{2}-\mathrm{sgn}(\beta-\alpha)\frac{\pi}{2},\nonumber \\
&&\theta=2\arctan\sqrt{\frac{1-\cos\left(\delta-\gamma\right)}{1-\cos\left(\beta-\alpha\right)}}=2\arctan\left[\sin\left(\frac{\delta-\gamma}{2}\right)\biggl/\sin\left(\frac{\beta-\gamma}{2}\right)\right]\hspace{.6cm}\mathrm{and}\nonumber \\
&&\phi=\frac{\gamma+\delta}{2}-\frac{\alpha+\beta}{2}-\frac{\pi}{2}\Bigl[\mathrm{sgn}(\delta-\gamma)-\mathrm{sgn}(\beta-\alpha)\Bigr],
\end{eqnarray}
where the first identity invoked in Eq.~(\ref{phase}) has been used again. Note that Eq.~(\ref{finally}) and Eq.~(\ref{parameters}) respectively match Eq.~(\ref{not_randomised}) and Eq.~(\ref{variables}) exactly, as we wanted to show.
\section{Calculation of the output probability density function of the transmitter}\label{distribution}
In this appendix, we shall use bold letters to denote random variables (RVs). The starting point is the definition of the output RVs $\boldsymbol{\phi}$, $\boldsymbol{\theta}$ and $\boldsymbol{I}$ of the PT, given in Eq.~(\ref{variables}):
\begin{eqnarray}\label{variables_2}
&&\boldsymbol{\phi}=\boldsymbol{\delta_{2}}+\frac{\boldsymbol{\delta_{1}}+\boldsymbol{\delta_{3}}}{2}-\frac{\pi}{2}\Bigl[\mathrm{sgn}(\boldsymbol{\delta_{3}})-\mathrm{sgn}(\boldsymbol{\delta_{1}})\Bigr],\nonumber \\
&&\boldsymbol{\theta}=2\arctan\left[\sin\left(\frac{\boldsymbol{\delta_{3}}}{2}\right)\biggl/\sin\left(\frac{\boldsymbol{\delta_{1}}}{2}\right)\right],\nonumber \\
&&\boldsymbol{I}=2\nu{}t\left[\sin^{2}\left(\frac{\boldsymbol{\delta_{1}}}{2}\right)+\sin^{2}\left(\frac{\boldsymbol{\delta_{3}}}{2}\right)\right].\nonumber \\
\end{eqnarray}

\textbf{Distribution and independence of $\boldsymbol{\phi}$}\\

In the first place, since $\boldsymbol{\delta_{1}}$, $\boldsymbol{\delta_{2}}$ and $\boldsymbol{\delta_{3}}$ are independent and uniformly distributed in $[0,2\pi)$, it trivially follows that $\boldsymbol{\phi}$ is uniformly distributed in $[0,2\pi)$ too. Namely, $f_{\boldsymbol{\phi}}(\phi)=1/2\pi$ for all $\phi\in[0,2\pi)$. We discuss the independence of $\boldsymbol{\phi}$ from the bivariate RV $(\boldsymbol{\theta},\boldsymbol{I})$ next. In fact, it suffices to notice that, according to Eq.~(\ref{variables_2}), $\boldsymbol{\phi}{\bigl|}_{(\boldsymbol{\delta_{1}},\boldsymbol{\delta_{3}})=\left(\delta_{1},\delta_{3}\right)}$ is of the form ``constant phase plus uniformly distributed phase", such that $f_{\boldsymbol{\phi}|(\boldsymbol{\delta_{1}},\boldsymbol{\delta_{3}})=\left(\delta_{1},\delta_{3}\right)}(\phi)=1/{2\pi}=f_{\boldsymbol{\phi}}(\phi)$ for all $\delta_{1}$, $\delta_{3}$. This independence between $\boldsymbol{\phi}$ and $(\boldsymbol{\delta_{1}},\boldsymbol{\delta_{3}})$ straightforwardly implies the independence of $\boldsymbol{\phi}$ from $(\boldsymbol{\theta},\boldsymbol{I})$.\\

\textbf{Distribution of $\boldsymbol{\theta}$ and $\boldsymbol{I}$}\\

Here, we compute the joint PDF of $\boldsymbol{\theta}$ and $\boldsymbol{I}$, $f_{\boldsymbol{\theta},\boldsymbol{I}}$, and recall that $f_{\boldsymbol{\phi},\boldsymbol{\theta},\boldsymbol{I}}=f_{\boldsymbol{\phi}}\times{}f_{\boldsymbol{\theta},\boldsymbol{I}}$ in virtue of the previous discussion.\\

The starting point is the joint PDF of the independent variables $\boldsymbol{\delta_{1}}$ and $\boldsymbol{\delta_{3}}$ ---given by $f_{\boldsymbol{\delta_{1}},\boldsymbol{\delta_{3}}}(\delta_{1},\delta_{3})=1/(2\pi)^{2}$ for all $(\delta_{1},\delta_{3})\in\mathcal{R}=[0,2\pi)\times[0,2\pi)$--- and the function $g$ that maps $(\boldsymbol{\delta_{1}},\boldsymbol{\delta_{3}})$ to $(\boldsymbol{\theta},\boldsymbol{I})$. For convenience, we shall deal with the dimensionless variable $\boldsymbol{y}=\boldsymbol{I}/2\nu{}t$, such that the relevant function $G$ (identical to $g$ up to a constant prefactor in the second component) reads
\begin{equation}\label{G}
G:\begin{cases}
\theta=\displaystyle{2\arctan\left[\sin\left(\frac{\delta_{3}}{2}\right)\biggl/\sin\left(\frac{\delta_{1}}{2}\right)\right]}\\
\\
y=\displaystyle{\sin^{2}\left(\frac{\delta_{1}}{2}\right)+\sin^{2}\left(\frac{\delta_{3}}{2}\right)}
\end{cases}
\end{equation}
Despite the non-injectiveness of $G$ in $\mathcal{R}$, $G$ is symmetric with respect to the axes $\delta_{1}=\pi$ and $\delta_{3}=\pi$, and its restriction $G|_{Q_{k}}$ to any of the four quadrants $Q_{k}$ of $\mathcal{R}$ defined by these axes is injective. Therefore, any $(\theta,y)$ in the interior of $G(\mathcal{R})$ accumulates the probability densities coming from all $G^{-1}(\theta,y)\cap{}Q_{k}$ (related with each other by reflections with respect to the axes), and in virtue of the bivariate transformation theorem it follows that
\begin{equation}\label{transformation_theorem}
f_{\boldsymbol{\theta},\boldsymbol{y}}(\theta,y)=\sum_{k=1}^{4}f_{\boldsymbol{\delta_{1}},\boldsymbol{\delta_{3}}}\left(G^{-1}(\theta,y)\cap{}Q_{k}\right)\abs{J_{G}\left(G^{-1}(\theta,y)\cap{}Q_{k}\right)}^{-1},
\end{equation}
where $J_{G}$ is the jacobian determinant of the $G$ function,
\begin{equation}\label{jacobian}
J_{G}(\delta_{1},\delta_{3})=\mathrm{det}
\begin{bmatrix}
\displaystyle{\frac{\partial\theta}{\partial\delta_{1}}} & \displaystyle{\frac{\partial\theta}{\partial\delta_{3}}} \\
\displaystyle{\frac{\partial y}{\partial \delta_{1}}} & \displaystyle{\frac{\partial y}{\partial \delta_{3}}} \\
\end{bmatrix}
=-\cos\left(\frac{\delta_{1}}{2}\right)\cos\left(\frac{\delta_{3}}{2}\right).
\end{equation}
We remark that the r.h.s. in Eq.~(\ref{jacobian}) follows from explicit calculation of the derivatives and the determinant. Notably, $\abs{J_{G}\left(\delta_{1},\delta_{3}\right)}$ is invariant under reflections with respect to $\delta_{1}=\pi$ and/or $\delta_{3}=\pi$, such that $\abs{J_{G}\left(G^{-1}(\theta,y)\cap{}Q_{k}\right)}$ in Eq.~(\ref{transformation_theorem}) takes the same value for all four contributions to the preimage of $(\theta,y)$. Since, in addition, $f_{\boldsymbol{\delta_{1}},\boldsymbol{\delta_{3}}}\left(\delta_{1},\delta_{3}\right)=1/(2\pi)^2$ for all pairs $(\delta_{1},\delta_{3})$, Eq.~(\ref{transformation_theorem}) simplifies as
\begin{equation}\label{transformation_theorem_2}
f_{\boldsymbol{\theta},\boldsymbol{y}}(\theta,y)=\pi^{-2}\abs{J_{G}\left(G^{-1}(\theta,y)\cap{}Q_{1}\right)}^{-1}.
\end{equation}
At the symmetry axes $\delta_{1}=\pi$ and $\delta_{3}=\pi$, which are necessarily mapped to the boundary of $G(\mathcal{R})$, $J_{G}\left(\delta_{1},\delta_{3}\right)$ vanishes. As a consequence, $f_{\boldsymbol{\theta},\boldsymbol{y}}$ is divergent in this frontier. Note, however, that this feature does not compromise the normalization of $f_{\boldsymbol{\theta},\boldsymbol{y}}$ or the finiteness of any physical quantity relevant to the problem.

All that remains is to write down the jacobian determinant of Eq.~(\ref{jacobian}) in terms of $\theta$ and $y$.
In order to do this, it suffices to notice that
\begin{equation}\label{trick_2}
\cos^{2}\left(\frac{\theta}{2}\right)=\frac{\displaystyle{\sin^{2}\left(\frac{\delta_{1}}{2}\right)}}{\displaystyle{\sin^{2}\left(\frac{\delta_{1}}{2}\right)+\sin^{2}\left(\frac{\delta_{3}}{2}\right)}}\hspace{.2cm}\mathrm{and}\hspace{.2cm}\sin^{2}\left(\frac{\theta}{2}\right)=\frac{\displaystyle{\sin^{2}\left(\frac{\delta_{3}}{2}\right)}}{\displaystyle{\sin^{2}\left(\frac{\delta_{1}}{2}\right)+\sin^{2}\left(\frac{\delta_{3}}{2}\right)}}
\end{equation}
in virtue of Eq.~(\ref{G}), such that
\begin{equation}
1-y{}\cos^{2}\left(\frac{\theta}{2}\right)=\cos^{2}\left(\frac{\delta_{1}}{2}\right)\hspace{.2cm}\mathrm{and}\hspace{.2cm}1-y{}\sin^{2}\left(\frac{\theta}{2}\right)=\cos^{2}\left(\frac{\delta_{3}}{2}\right).
\end{equation}
Substituting these two relations in Eq.~(\ref{jacobian}) and plugging the result in Eq.~(\ref{transformation_theorem_2}) yields
\begin{equation}
f_{\boldsymbol{\theta},\boldsymbol{y}}(\theta,y)=\frac{1}{\pi^{2}\sqrt{1-y{}\displaystyle{\cos^{2}\left(\frac{\theta}{2}\right)}}\sqrt{1-y{}\displaystyle{\sin^{2}\left(\frac{\theta}{2}\right)}}}.
\end{equation}

To finish with, let us explicitly identify the domain of $f_{\boldsymbol{\theta},\boldsymbol{y}}(\theta,y)$, given by the image of the domain $\mathcal{R}$ via $G$. In virtue of the symmetry of $G$, $G(\mathcal{R})=G(Q_{k})$ for all $k$, and thus it suffices to show that $G$ maps, say, the quadrant $Q_{1}=[0,\pi]\times[0,\pi]$ to the region
\begin{equation}\label{region}
G(Q_{1})=\left\{\theta\in[0,\pi],y\in\left[0,y_{\mathrm{max},\theta}\right]\right\},
\end{equation}
where $y_{\mathrm{max},\theta}=\min\left\{1/\cos^{2}\left(\theta/2\right),1/\sin^{2}\left(\theta/2\right)\right\}$. For this purpose, we identify the image of the boundary of $Q_{1}$ via $G$, which certainly defines the
boundary of $G(Q_{1})$. If, for instance, we label the sides of the rectangle $Q_{1}$ as $L_{1}=\{\delta_{1}\in[0,\pi],\delta_{3}=0\}$, $L_{2}=\{\delta_{1}\in[0,\pi],\delta_{3}=\pi\}$, $L_{3}=\{\delta_{1}=0,\delta_{3}\in[0,\pi]\}$ and $L_{4}=\{\delta_{1}=\pi,\delta_{3}\in[0,\pi]\}$, one can readily show that $L_{1}$ contributes with the border $G(L_{1})=\{\theta=0,y\in[0,1]\}$, $L_{2}$ with the border $G(L_{2})=\{\theta\in[\pi/2,\pi],y=\sin\left(\theta/2\right)^{-2}\}$, $L_{3}$ with the border $G(L_{3})=\{\theta=\pi,y\in[0,1]\}$ and $L_{4}$ with the border $G(L_{4})=\{\theta\in[0,\pi/2],y=\cos\left(\theta/2\right)^{-2}\}$. These four borders (together with the defining constraint $y
\geq{}0$) shape the boundary of the region $G(Q_{1})=G(\mathcal{R})$ defined in Eq.~(\ref{region}).

Needless to say, $f_{\boldsymbol{\theta},\boldsymbol{I}}(\theta,I)$ follows trivially from $f_{\boldsymbol{\theta},\boldsymbol{y}}(\theta,y)$ and the fact that $\boldsymbol{y}=\boldsymbol{I}/2\nu{}t$, leading to Eq.~(\ref{PDF_2}) in the main text (where the border $\left\{\theta\in[0,\pi],y=y_{\mathrm{max},\theta}\right\}$ is excluded because of the divergence of $f_{\boldsymbol{\theta},\boldsymbol{I}}$).
\section{Trace distance argument}\label{TD}
Let $\rho$ and $\sigma$ be two density matrices of a quantum system of dimension $d$. The TD between them is defined as $D(\rho,\sigma)=\frac{1}{2}\Tr\left[\sqrt{\left(\rho-\sigma\right)^{2}}\right]$, and the TD argument states that $D(\rho,\sigma)=\max\left\{\mathrm{Tr}\left[\hat{O}(\rho-\sigma)\right]\right\}$, where the maximization is taken over all positive operators $\hat{O}\leq{I}$~\cite{Nielsen}.\\

Notably, from the definition of the TD it follows that $D(\rho,\sigma)=\sum_{i=1}^{d}\abs{\lambda_{i}}$, where the $\lambda_{i}$ are the eigenvalues of $\rho-\sigma$.
\section{Numerical evaluation of the trace distance constraints}\label{representation}
In order to evaluate the TD constraints of Sec.~\ref{decoy-state}, we express the Fock states $\sigma_{x,j,n}$ (defined in Eq.~(\ref{Fock})) in a computational basis. For this purpose, we work with the unnormalized states
\begin{equation} \tilde{\sigma}_{x,j,n}=\left\langle\frac{e^{-I}{I}^{n}}{n!}\ketbra{n}{n}_{\theta,\phi}\right\rangle_{\Omega_{x,j}}
\end{equation}
instead, and recall that $\sigma_{x,j,n}=\tilde{\sigma}_{x,j,n}/\Tr\left[\tilde{\sigma}_{x,j,n}\right]$ with $\Tr\left[\tilde{\sigma}_{x,j,n}\right]=\left\langle{e^{-I}{I}^{n}/{n!}}\right\rangle_{\Omega_{x,j}}$. The preferred basis that we use here is the one induced by the creation operators $a^{\dagger}_{\rm R}$ and $a^{\dagger}_{\rm L}$ presented in Sec.~\ref{Characterization}:
\begin{equation}
\mathcal{B}_{n}=\left\{\ket{n-k,k}=\frac{a^{\dagger n-k}_{\rm R}a^{\dagger k}_{\rm L}}{\sqrt{(n-k)!k!}}\ket{\rm vac},\ k=0,\ldots,n\right\}.
\end{equation}
Notably, $\mathcal{B}_{n}$ is an orthonormal basis of the Hilbert space $\mathcal{H}_{n}$ of $n$ indistinguishable photons distributed across two modes, such that $\dim{\mathcal{H}_{n}}=n+1$. In particular, the states $\ketbra{n}{n}_{\theta,\phi}$ (defined in Eq.~(\ref{fock_theta_phi})) trivially decompose as
\begin{equation}\label{Fock_decomposition}
\ketbra{n}{n}_{\theta,\phi}=\sum_{k=0}^{n}\sum_{l=0}^{n}\sqrt{\binom{n}{k}\binom{n}{l}}e^{i(k-l)\phi}\cos^{2n-(k+l)}\left(\frac{\theta}{2}\right)\sin^{k+l}\left(\frac{\theta}{2}\right)\ketbra{n-k,k}{n-l,l}
\end{equation}
in virtue of Newton's binomial formula. Now, in contrast to the states $\ket{n}_{\theta,\phi}$, the basis elements $\ket{n-k,k}$ are independent of $\theta$ and $\phi$, thereby allowing us to proceed with the angular integrals in
\begin{equation}\label{explicit}
\left\langle\frac{e^{-I}{I}^{n}}{n!}\ketbra{n}{n}_{\theta,\phi}\right\rangle_{\Omega_{x,j}}=\frac{1}{2\pi}\int_{x-\Delta\phi}^{x+\Delta\phi}d\phi\int_{\frac{\pi}{2}-\Delta\theta}^{\frac{\pi}{2}+\Delta\theta}d\theta\int_{I_{j}}dI{}f_{\boldsymbol{\theta},\boldsymbol{I}}(\theta,I)\frac{e^{-I}{I}^{n}}{n!}\ketbra{n}{n}_{\theta,\phi}.
\end{equation}
Specifically, the relevant azimuthal integral in Eq.~(\ref{explicit}) is given by
\begin{equation}\label{azimuthal}
\frac{1}{2\pi}\int_{x-\Delta\phi}^{x+\Delta\phi}d\phi{}e^{i(k-l)\phi}=\begin{cases}
\displaystyle{\frac{\Delta\phi}{\pi}} & \mathrm{if}\hspace{.3cm}k=l, \\
\\
\displaystyle{\frac{\sin\left[(k-l)\Delta\phi\right]}{(k-l)\pi}}\displaystyle{e^{i(k-l)x}} & \mathrm{if}\hspace{.3cm}k\neq{}l,
\end{cases}
\end{equation}
such that
\begin{eqnarray}\label{Fock_decomposition_2}
&&\frac{1}{2\pi}\int_{x-\Delta\phi}^{x+\Delta\phi}d\phi{}\ketbra{n}{n}_{\theta,\phi}=\sum_{k=0}^{n}\binom{n}{k}\frac{\Delta\phi}{\pi}\cos^{2(n-k)}\left(\frac{\theta}{2}\right)\sin^{2k}\left(\frac{\theta}{2}\right)\ketbra{n-k,k}{n-k,k}\nonumber \\
&&+\sum_{k=1}^{n}\sum_{l<k}\sqrt{\binom{n}{k}\binom{n}{l}}\frac{\sin\left[(k-l)\Delta\phi\right]}{(k-l)\pi}\cos^{2n-(k+l)}\left(\frac{\theta}{2}\right)\sin^{k+l}\left(\frac{\theta}{2}\right)\left[e^{i(k-l)x}\ketbra{n-k,k}{n-l,l}+e^{-i(k-l)x}\ketbra{n-l,l}{n-k,k}\right].\nonumber \\
&&
\end{eqnarray}
Note that this result is obtained by simply splitting $\ketbra{n}{n}_{\theta,\phi}$ in Eq.~(\ref{Fock_decomposition}) into diagonal and off-diagonal terms (in the $\mathcal{B}_{n}$ basis), and then using Eq.~(\ref{azimuthal}) for the integration in $\phi$. Finally, plugging Eq.~(\ref{Fock_decomposition_2}) into Eq.~(\ref{explicit}) yields
\begin{eqnarray}\label{matlab_input}
&&\tilde{\sigma}_{x,j,n}=\sum_{k=0}^{n}\binom{n}{k}\frac{\Delta\phi}{\pi}\left\{\int_{\frac{\pi}{2}-\Delta\theta}^{\frac{\pi}{2}+\Delta\theta}d\theta\cos^{2(n-k)}\left(\frac{\theta}{2}\right)\sin^{2k}\left(\frac{\theta}{2}\right)\int_{I_{j}}dI{}f_{\boldsymbol{\theta},\boldsymbol{I}}(\theta,I)\frac{e^{-I}{I}^{n}}{n!}\right\}\times\ketbra{n-k,k}{n-k,k}\nonumber \\
&&+\sum_{k=1}^{n}\sum_{l<k}\sqrt{\binom{n}{k}\binom{n}{l}}\ \frac{\sin\left[(k-l)\Delta\phi\right]}{(k-l)\pi}\left\{\int_{\frac{\pi}{2}-\Delta\theta}^{\frac{\pi}{2}+\Delta\theta}d\theta\cos^{2n-(k+l)}\left(\frac{\theta}{2}\right)\sin^{k+l}\left(\frac{\theta}{2}\right)\int_{I_{j}}dI{}f_{\boldsymbol{\theta},\boldsymbol{I}}(\theta,I)\frac{e^{-I}{I}^{n}}{n!}\right\}\times\nonumber \\
&&\left[e^{i(k-l)x}\ketbra{n-k,k}{n-l,l}+e^{-i(k-l)x}\ketbra{n-l,l}{n-k,k}\right].
\end{eqnarray}

Now, we make use of the canonical isomorphism:
\begin{equation}
\ket{n,0}\rightarrow{}[1\hspace{.05cm}0\ldots\hspace{.05cm}0]^{t},\hspace{.1cm}\ket{n-1,1}\rightarrow{}[0\hspace{.05cm}1\ldots\hspace{.05cm}0]^{t},\hspace{.05cm}\ldots\hspace{.1cm},\hspace{.1cm}\ket{0,n}\rightarrow{}[0\hspace{.05cm}\ldots\hspace{.05cm}0\hspace{.05cm}1]^{t}.
\end{equation}
This provides a natural matrix representation of the $\sigma_{x,j,n}$, where the $(r,s)$-th entry is given by
\begin{equation}
\bra{n-r+1,r-1}\sigma_{x,j,n}\ket{n-s+1,s-1}
\end{equation}
for $r,s=1,\ldots,n+1$. These matrices can be written down in any scientific computing tool for the numerical calculation of the TD via the eigenvalues, as indicated in Appendix~\ref{TD}.
\section{Derivation of Eq.~(\ref{test})}\label{EB}
The goal is to show that
\begin{equation}\label{goal}
\frac{1}{\sqrt{2}}\Bigl(\ket{0}_{\mathrm{A'}}\ket{\Psi_{0,\mathrm{s},1}}_{\rm AB}+\ket{\pi}_{\mathrm{A'}}\ket{\Psi_{\pi,\mathrm{s},1}}_{\rm AB}\Bigr)=\frac{1}{\sqrt{2}}\left(\ket{\frac{\pi}{2}}_{\mathrm{A'}}\ket{\Psi_{-\frac{\pi}{2},\mathrm{s},1}}_{\rm AB}+\ket{-\frac{\pi}{2}}_{\mathrm{A'}}\ket{\Psi_{\frac{\pi}{2},\mathrm{s},1}}_{\rm AB}\right).
\end{equation}
From the RL Bloch sphere notation, $\ket{y}_{\rm A'}=\frac{1}{\sqrt{2}}\left(\ket{\rm R}_{\rm A'}+e^{iy}\ket{\rm L}_{\rm A'}\right)$, one can readily show that
\begin{equation}\label{A}
\ket{0}_{\rm A'}=\frac{e^{-i\frac{\pi}{4}}}{\sqrt{2}}\left(\ket{\frac{\pi}{2}}_{\mathrm{A'}}+i\ket{-\frac{\pi}{2}}_{\mathrm{A'}}\right)\hspace{.2cm}\mathrm{and}\hspace{.2cm}\ket{\pi}_{\rm A'}=\frac{e^{i\frac{\pi}{4}}}{\sqrt{2}}\left(\ket{\frac{\pi}{2}}_{\mathrm{A'}}-i\ket{-\frac{\pi}{2}}_{\mathrm{A'}}\right),
\end{equation}
where we keep the global phases for clarity. Plugging these relations into the l.h.s. of Eq.~(\ref{goal}) and reordering yields
\begin{eqnarray}\label{intermediate}
&&\frac{1}{\sqrt{2}}\Bigl(\ket{0}_{\mathrm{A'}}\ket{\Psi_{0,\mathrm{s},1}}_{\rm AB}+\ket{\pi}_{\mathrm{A'}}\ket{\Psi_{\pi,\mathrm{s},1}}_{\rm AB}\Bigr)=\nonumber \\
&&=\frac{1}{\sqrt{2}}\left(\ket{\frac{\pi}{2}}_{\mathrm{A'}}\frac{e^{-i\frac{\pi}{4}}\ket{\Psi_{0,\mathrm{s},1}}_{\rm AB}+e^{i\frac{\pi}{4}}\ket{\Psi_{\pi,\mathrm{s},1}}_{\rm AB}}{\sqrt{2}}+\ket{-\frac{\pi}{2}}_{\mathrm{A'}}\frac{e^{i\frac{\pi}{4}}\ket{\Psi_{0,\mathrm{s},1}}_{\rm AB}+e^{-i\frac{\pi}{4}}\ket{\Psi_{\pi,\mathrm{s},1}}_{\rm AB}}{\sqrt{2}}\right).
\end{eqnarray}
Similarly, from the definition of $\ket{\Psi_{x,\mathrm{s},1}}_{\rm AB}$ (given in Eq.~(\ref{shield})) and the RL Bloch sphere notation for system B, $\ket{x}_{\rm B}=\frac{1}{\sqrt{2}}\left(\ket{\rm R}_{\rm B}+e^{ix}\ket{\rm L}_{\rm B}\right)$, one can easily show that
\begin{equation}\label{B}
\frac{e^{-i\frac{\pi}{4}}\ket{\Psi_{0,\mathrm{s},1}}_{\rm AB}+e^{i\frac{\pi}{4}}\ket{\Psi_{\pi,\mathrm{s},1}}_{\rm AB}}{\sqrt{2}}=\ket{\Psi_{-\frac{\pi}{2},\mathrm{s},1}}_{\rm AB}\hspace{.2cm}\mathrm{and}\hspace{.2cm}\frac{e^{i\frac{\pi}{4}}\ket{\Psi_{0,\mathrm{s},1}}_{\rm AB}+e^{-i\frac{\pi}{4}}\ket{\Psi_{\pi,\mathrm{s},1}}_{\rm AB}}{\sqrt{2}}=\ket{\Psi_{\frac{\pi}{2},\mathrm{s},1}}_{\rm AB}.
\end{equation}
Plugging these relations into Eq.~(\ref{intermediate}), Eq.~(\ref{goal}) follows.
\section{Channel model}\label{channel_model}
\begin{figure}[!htbp]
	\centering 
	\includegraphics[width=5cm,height=2cm]{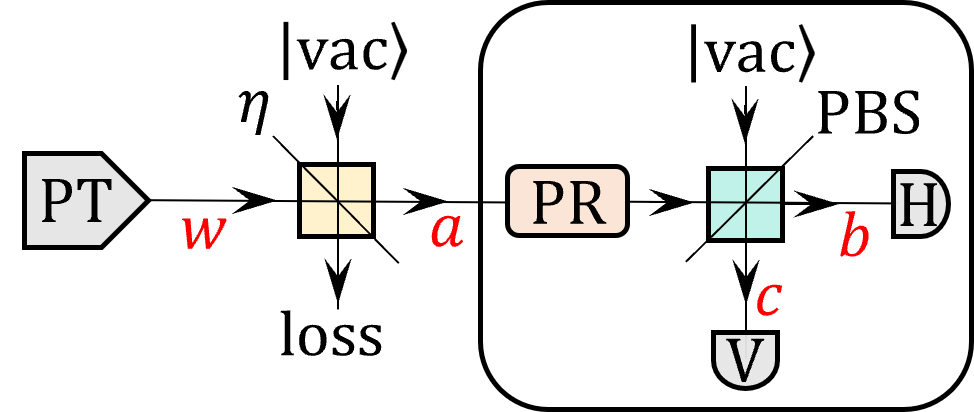}\\
	\caption{Schematic of the channel model. The quantum states generated in the PT undergo (1) a lossy quantum channel and (2) a quantum measurement at Bob's lab. The measurement is modelled with a polarization rotator (PR) that selects Bob's basis setting, followed by a PBS that splits orthogonal polarizations in the $Z$ basis (say, H and V). A noisy-and-lossy threshold detector is located in each output port of the PBS and we assume that both detectors are equal. Notably, channel and detector loss are jointly accounted for with a BS of effective transmittance $\eta=\eta_{\rm ch}\eta_{\rm det}$, where $\eta_{\rm ch}$ ($\eta_{\rm det}$) stands for the channel transmittance (detection efficiency). Also, the relevant spatial modes are indicated in red.}
	\label{channel}
\end{figure}

Here, we present the channel model that we use for the simulations. As shown in Sec.~\ref{Characterization}, the PT generates the phase-randomised WCP
\begin{equation}
\rho_{w}^{I,\theta,\phi}=\sum_{n=0}^{\infty}\frac{e^{-I}{I}^{n}}{n!}\ketbra{n}{n}_{\theta,\phi}
\end{equation}
in, say spatial mode $w$, with a known probability density function $f_{\boldsymbol{\phi},\boldsymbol{\theta},\boldsymbol{I}}$. This state can also be written in the form
\begin{equation}
\rho_{w}^{I,\theta,\phi}=\frac{1}{2\pi}\int_{0}^{2\pi}d\psi\ketbra{\sqrt{I}\ e^{i\psi}}{\sqrt{I}\ e^{i\psi}}_{w,\theta,\phi},
\end{equation}
such that one can apply the channel model to the pure state $\ket{\sqrt{I}\ e^{i\psi}}_{w,\theta,\phi}$ first and proceed with the phase-averaging later on. This is what we do next. The process that $\ket{\sqrt{I}\ e^{i\psi}}_{w,\theta,\phi}$ undergoes is illustrated in Fig.~\ref{channel}.\\

The BS transformation simply maps $\ket{\sqrt{I}\ e^{i\psi}}_{w,\theta,\phi}$ (in spatial mode $w$ in the figure) to $\ket{\sqrt{I\eta}\ e^{i\psi}}_{a,\theta,\phi}$ (in spatial mode $a$ in the figure). Now, in order to describe the measurement statistics of the basis-matched events, it suffices to contemplate one basis. For instance, let us consider that the generated state $\ket{\sqrt{I}\ e^{i\psi}}_{w,\theta,\phi}$ lies in the $Z$ basis acceptance region ---meaning that $(\phi,\theta,I)\in\Omega_{j}^{Z}$ for some $j\in\{\mathrm{s,d,v}\}$--- and Bob selects the $Z$ basis for his measurement too ---meaning that the polarization rotator (PR) does not alter the incident polarization---. Aiming to incorporate the action of the PBS, we recall that $\ket{\sqrt{I\eta}\ e^{i\psi}}_{a,\theta,\phi}=\exp\bigl\{\sqrt{I\eta}e^{i\psi}{}a_{\theta,\phi}^{\dagger}-\sqrt{I\eta}e^{-i\psi}{}a_{\theta,\phi}\bigr\}\ket{\rm vac}$ with $a^{\dagger}_{\theta,\phi}=\cos\left({\theta}/{2}\right)a^{\dagger}_{\rm R}+e^{i\phi}\sin\left({\theta}/{2}\right)a^{\dagger}_{\rm L}$, and rewrite this state in terms of the creation operators associated to the $Z$ basis, defined as $a_{\rm H}^{\dagger}=a^{\dagger}_{\frac{\pi}{2},0}$, $a_{\rm V}^{\dagger}=a^{\dagger}_{\frac{\pi}{2},\pi}$. This yields
\begin{equation}
\ket{\sqrt{I\eta}\ e^{i\psi}}_{a,\theta,\phi}=\exp\left\{\sqrt{\frac{I\eta}{2}}e^{i\psi}\left[\Bigl(\cos\left({\theta}/{2}\right)+e^{i\phi}\sin\left({\theta}/{2}\right)\Bigr)a_{\rm H}^{\dagger}+\Bigl(\cos\left({\theta}/{2}\right)-e^{i\phi}\sin\left({\theta}/{2}\right)\Bigr)a_{\rm V}^{\dagger}\right]-\widehat{\mathrm{h.c.}}\right\}\ket{\rm vac}.
\end{equation}
The transformation of the PBS reads $\left\{a_{\rm H}^{\dagger}\rightarrow{}b_{\rm H}^{\dagger},\ a_{\rm V}^{\dagger}\rightarrow{}c_{\rm V}^{\dagger}\right\}$, which maps $\ket{\sqrt{I\eta}\ e^{i\psi}}_{a,\theta,\phi}$ to
\begin{eqnarray}\label{PBS_transformation}
&&\ket{\sqrt{\frac{I\eta}{2}}e^{i\psi}\Bigl(\cos\left({\theta}/{2}\right)+e^{i\phi}\sin\left({\theta}/{2}\right)\Bigr)}_{b,\frac{\pi}{2},0}\ket{\sqrt{\frac{I\eta}{2}}e^{i\psi}\Bigl(\cos\left({\theta}/{2}\right)-e^{i\phi}\sin\left({\theta}/{2}\right)\Bigr)}_{c,\frac{\pi}{2},\pi}=\nonumber \\
&&\ket{\sqrt{\frac{I\eta\left(1+\sin\theta\cos\phi\right)}{2}}e^{i\Gamma_{+}}}_{b,\frac{\pi}{2},0}\ket{\sqrt{\frac{I\eta\left(1-\sin\theta\cos\phi\right)}{2}}e^{i\Gamma_{-}}}_{c,\frac{\pi}{2},\pi},
\end{eqnarray}
for known phases $\Gamma_{+}$ and $\Gamma_{-}$ irrelevant to the discussion.

Coming next, the measurement implemented by each photo-detector is described by a POVM with elements $\{\hat{E}_{\mathrm{no\hspace{.05cm}click}}=(1-p_{\rm d})\ketbra{0}{0},\hat{E}_{\mathrm{click}}=\mathds{1}-\hat{E}_{\mathrm{no\hspace{.05cm}click}}\}$, where $p_{\rm d}$ stands for the dark count probability and $\mathds{1}$ stands for the identity operator (note that detector loss has already been accounted for in the channel). Denoting the first (second) ket in the r.h.s. of Eq.~(\ref{PBS_transformation}) by $\ket{\rm B}$ ($\ket{\rm C}$), it follows that detectors ``H" and ``V" in Fig.~\ref{channel} record a click with independent probabilities $p_{\rm H}=\Tr\bigl[\hat{E}_{\mathrm{click}}\ketbra{B}{B}\bigr]$ and $p_{\rm V}=\Tr\bigl[\hat{E}_{\mathrm{click}}\ketbra{C}{C}\bigr]$, such that
\begin{equation}
p_{\rm H}=1-(1-p_{\rm d})e^{\displaystyle{-\frac{I\eta\left(1+\sin\theta\cos\phi\right)}{2}}}\hspace{.3cm}\mathrm{and}\hspace{.3cm}p_{\rm V}=1-(1-p_{\rm d})e^{\displaystyle{-\frac{I\eta\left(1-\sin\theta\cos\phi\right)}{2}}}.
\end{equation}
These probabilities do not depend on the phase $\psi$, which implies that they remain unchanged after phase averaging. In short, $\rho_{w}^{I,\theta,\phi}$ triggers a click when measured in the $Z$ basis with overall probability
\begin{equation}\label{yield_rho}
p\left(\mathrm{click}\bigl|\rho_{w}^{I,\theta,\phi},Z\right)=1-p\left(\mathrm{no\hspace{.05cm}click}\bigl|\rho_{w}^{I,\theta,\phi},Z\right)=1-(1-p_{\rm H})(1-p_{\rm V})=1-(1-p_{\rm d})^{2}e^{\displaystyle{-I\eta}}.
\end{equation}

Notably, Eq.~(\ref{yield_rho}) does not use the fact that $\rho_{w}^{I,\theta,\phi}$ belongs to a $Z$ basis acceptance region in any way. This consideration only matters for the calculation of the error probability $p\left(\mathrm{err}\bigl|\rho_{w}^{I,\theta,\phi},Z\right)$, which we do next. For this purpose, we further assume that $\rho_{w}^{I,\theta,\phi}$ belongs to an acceptance region associated to the horizontal polarization. This amounts to saying that $\phi\in[x-\Delta\phi,x+\Delta\phi]$ with $x=0$, or equivalently that $(\phi,\theta,I)\in\Omega_{0,j}$ for some $j\in\{\mathrm{s,d,v}\}$. In this case, a bit error occurs if the outcome ``V" is recorded. Considering, as usual, that double-clicks are randomly assigned to a specific outcome, this means that $p\left(\mathrm{err}\bigl|\rho_{w}^{I,\theta,\phi},Z\right)=p_{\rm V}(1-p_{\rm H})+p_{\rm V}p_{\rm H}/2$ for all $(\phi,\theta,I)\in\Omega_{0,j}$, which is easily taken to the form
\begin{equation}\label{error_probability_rho}
p\left(\mathrm{err}\bigl|\rho_{w}^{I,\theta,\phi},Z\right)=\frac{1}{2}\left[1-(1-p_{\rm d})^{2}e^{\displaystyle{-I\eta}}\right]-\frac{1}{2}\left(1-p_{\rm d}\right)\left[e^{\displaystyle{-\frac{I\eta\left(1-\sin\theta\cos\phi\right)}{2}}}-e^{\displaystyle{-\frac{I\eta\left(1+\sin\theta\cos\phi\right)}{2}}}\right].
\end{equation}
Finally, given Eq.~(\ref{yield_rho}) and Eq.~(\ref{error_probability_rho}), it is straightforward to compute $Q^{Z}_{j}=p\left(\mathrm{click}|\sigma_{j}^{Z},Z\right)$ and $E^{Z}_{j}=p\left(\mathrm{err}|\sigma_{j}^{Z},Z\right)$. To be precise, since $\sigma_{j}^{Z}=\left\langle\rho_{w}^{I,\theta,\phi}\right\rangle_{\Omega_{j}^{Z}}/{\left\langle{1}\right\rangle_{\Omega_{j}^{Z}}}$, it follows that
\begin{equation}
Q^{Z}_{j}=\frac{\left\langle{}p\left(\mathrm{click}\bigl|\rho_{w}^{I,\theta,\phi},Z\right)\right\rangle_{\Omega_{j}^{Z}}}{\left\langle{1}\right\rangle_{\Omega_{j}^{Z}}}=1-(1-p_{\rm d})^{2}\frac{\left\langle{e^{\displaystyle{-I\eta}}}\right\rangle_{\Omega_{j}^{Z}}}{\left\langle{1}\right\rangle_{\Omega_{j}^{Z}}}
\end{equation}
On the other hand, $E^{Z}_{j}=\bigl(p\left(\mathrm{err}|\sigma_{0,j},Z\right)+p\left(\mathrm{err}|\sigma_{\pi,j},Z\right)\bigr)/2$, and both $\sigma_{0,j}$ and $\sigma_{\pi,j}$ are equally likely to trigger an error for symmetry reasons within our channel model. Therefore, $E^{Z}_{j}=p\left(\mathrm{err}|\sigma_{0,j},Z\right)$, and thus
\begin{eqnarray}
&&E^{Z}_{j}=\frac{\left\langle{}p\left(\mathrm{err}\bigl|\rho_{w}^{I,\theta,\phi},Z\right)\right\rangle_{\Omega_{0,j}}}{\left\langle{1}\right\rangle_{\Omega_{0,j}}}=\nonumber \\
&&\frac{1}{2}\left[1-(1-p_{\rm d})^{2}\frac{\left\langle{e^{\displaystyle{-I\eta}}}\right\rangle_{\Omega_{0,j}}}{\left\langle{1}\right\rangle_{\Omega_{0,j}}}\right]-\frac{1}{2}\left(1-p_{\rm d}\right)\left[\frac{\left\langle{e^{\displaystyle{-\frac{I\eta\left(1-\sin\theta\cos\phi\right)}{2}}}}\right\rangle_{\Omega_{0,j}}}{\left\langle{1}\right\rangle_{\Omega_{0,j}}}-\frac{\left\langle{e^{\displaystyle{-\frac{I\eta\left(1+\sin\theta\cos\phi\right)}{2}}}}\right\rangle_{\Omega_{0,j}}}{\left\langle{1}\right\rangle_{\Omega_{0,j}}}\right].\nonumber \\
&&
\end{eqnarray}

\section{References}
\begin{thebibliography}{44}
%
\bibitem{PortmannRenner}
Portmann, C. \& Renner, R. Security in quantum cryptography. \textit{Rev. Mod. Phys.} \textbf{94}, 025008 (2022).
%
\bibitem{Curty}
Lo, H.-K., Curty, M. \& Tamaki, K. Secure quantum key distribution. \textit{Nat. Photonics} \textbf{8}, 595 (2014).
%
\bibitem{Feihu}
Xu, F., Ma, X., Zhang, Q., Lo, H.-K. \& Pan, J.-W. Secure quantum key distribution with realistic devices. \textit{Rev. Mod. Phys.} \textbf{92}, 025002 (2020).
%
\bibitem{Vakhitov}
Vakhitov, A., Makarov, V. \& Hjelme, D. R. Large pulse attack as a method of conventional optical eavesdropping in quantum cryptography. \textit{J. Mod. Opt.} \textbf{48}, 2023-2038 (2001).
%
%
\bibitem{Gisin}
Gisin, N., Fasel, S., Kraus, B., Zbinden, H. \& Ribordy, G. Trojan-horse attacks on quantum-key-distribution systems. \textit{Phys. Rev. A} \textbf{73}, 022320 (2006).
%
\bibitem{Jain1}
Jain, N., Stiller, B., Khan, I., Makarov, V., Marquardt, C. \& Leuchs, G. Risk analysis of Trojan-horse attacks on practical quantum key distribution systems. \textit{IEEE J. Sel. Top. Quantum Electron.} \textbf{21}, 168-177 (2014).
%
\bibitem{Jain2}
Jain, N., Anisimova, E., Khan, I., Makarov, V., Marquardt, C. \& Leuchs, G. Trojan-horse attacks threaten the security of practical quantum cryptography. \textit{New J. Phys.} \textbf{16}, 123030 (2014).
%
\bibitem{Sajeed}
Sajeed, S., Minshull, C., Jain, N. \& Makarov, V. Invisible Trojan-horse attack. \textit{Sci. Rep.} \textbf{7}, 1-7 (2017).
%
\bibitem{Lucamarini}
Lucamarini, M., Choi, I., Ward, M. B., Dynes, J. F., Yuan, Z. L. \& Shields, A. J. Practical security bounds against the trojan-horse attack in quantum key distribution. \textit{Phys. Rev. X} \textbf{5}, 031030 (2015).
%
\bibitem{Tamaki}
Tamaki, K., Curty, M. \& Lucamarini, M. Decoy-state quantum key distribution with a leaky source. \textit{New J. Phys.} \textbf{18}, 065008 (2016).
%
\bibitem{Weilong}
Wang, W., Tamaki, K. \& Curty, M. Finite-key security analysis for quantum key distribution with leaky sources. \textit{New J. Phys.} \textbf{20}, 083027 (2018).
%
\bibitem{Navs}
Navarrete, Á. \& Curty, M. Improved finite-key security analysis of quantum key distribution against Trojan-horse attacks. \textit{Quantum Sci. Technol.} \textbf{7}, 035021 (2022).
%
\bibitem{Liu}
Liu, B. \textit{et al}. Fully passive entanglement based quantum key distribution scheme. \textit{arXiv preprint} arXiv:2111.03211 (2021).
%
\bibitem{Paraiso}
Paraïso, T. K. \textit{et al}. A modulator-free quantum key distribution transmitter chip. \textit{npj Quantum Inf.} \textbf{5}, 1-6 (2019).
%
%
%
%
\bibitem{PDC_1}
Mauerer, W. \& Silberhorn, C. Quantum key distribution with passive decoy state selection. \textit{Phys. Rev. A} \textbf{75}, 050305 (2007).
%
\bibitem{PDC_2}
Wang, Q., Wang, X. B., Björk, G. \& Karlsson, A. Improved practical decoy state method in quantum key distribution with parametric down-conversion source. \textit{EPL} \textbf{79}, 40001 (2007).
%
\bibitem{PDC_3}
Adachi, Y., Yamamoto, T., Koashi, M. \& Imoto, N. Simple and efficient quantum key distribution with parametric down-conversion. \textit{Phys. Rev. Lett.} \textbf{99}, 180503 (2007).
%
\bibitem{PDC_4}
Ma, X. \& Lo, H.-K. Quantum key distribution with triggering parametric down-conversion sources. \textit{New J. Phys.} \textbf{10}, 073018 (2008).
%
\bibitem{PDC_5}
Adachi, Y., Yamamoto, T., Koashi, M. \& Imoto, N. Boosting up quantum key distribution by learning statistics of practical single-photon sources. \textit{New J. Phys.} \textbf{11}, 113033 (2009).
%
%
\bibitem{PDC_6}
Wang, Q., Zhang, C. H. \& Wang, X. B. Scheme for realizing passive quantum key distribution with heralded single-photon sources. \textit{Phys. Rev. A} \textbf{93}, 032312 (2016).
%
\bibitem{WCP_1}
Curty, M., Moroder, T., Ma, X. \& Lütkenhaus, N. Non-Poissonian statistics from Poissonian light sources with application to passive decoy state quantum key distribution. \textit{Opt. Lett.} \textbf{34}, 3238-3240 (2009).
%
\bibitem{WCP_2}
Curty, M., Ma, X., Qi, B. \& Moroder, T. Passive decoy-state quantum key distribution with practical light sources. \textit{Phys. Rev. A} \textbf{81}, 022310 (2010).
%
\bibitem{WCP_3}
Li, Y., Bao, W. S., Li, H. W., Zhou, C. \& Wang, Y. Passive decoy-state quantum key distribution using weak coherent pulses with intensity fluctuations. \textit{Phys. Rev. A} \textbf{89}, 032329 (2014).
%
\bibitem{WCP_4}
Shan, Y. Z., Sun, S. H., Ma, X. C., Jiang, M. S., Zhou, Y. L. \& Liang, L. M. Measurement-device-independent quantum key distribution with a passive decoy-state method. \textit{Phys. Rev. A} \textbf{90}, 042334 (2014).
%
\bibitem{experiment_1}
Zhang, Y. \textit{et al}. Practical non-Poissonian light source for passive decoy state quantum key distribution. \textit{Optics Lett.} \textbf{35}, 3393-3395 (2010).
%
\bibitem{experiment_2}
Zhang, Y. \textit{et al}. Experimental demonstration of passive decoy state quantum key distribution. \textit{Chinese Phys. B} \textbf{21}, 100307 (2012).
%
\bibitem{experiment_4}
Krapick, S., Stefszky, M. S., Jachura, M., Brecht, B., Avenhaus, M. \& Silberhorn, C. Bright integrated photon-pair source for practical passive decoy-state quantum key distribution. \textit{Phys. Rev. A} \textbf{89}, 012329 (2014).
%
\bibitem{experiment_3}
Sun, Q. C. \textit{et al}. Experimental passive decoy-state quantum key distribution. \textit{Laser Phys. Lett.} \textbf{11}, 085202 (2014).
%
\bibitem{experiment_5}
Guan, J. Y. \textit{et al.} Experimental passive round-robin differential phase-shift quantum key distribution. \textit{Phys. Rev. Lett.} \textbf{114}, 180502 (2015).
%
\bibitem{experiment_6}
Sun, S. H., Tang, G. Z., Li, C. Y. \& Liang, L. M. Experimental demonstration of passive-decoy-state quantum key distribution with two independent lasers. \textit{Phys. Rev. A} \textbf{94}, 032324 (2016).
%
\bibitem{passive_BB84}
Curty, M., Ma, X., Lo, H. K. \& Lütkenhaus, N. Passive sources for the Bennett-Brassard 1984 quantum-key-distribution protocol with practical signals. \textit{Phys. Rev. A} \textbf{82}, 052325 (2010).
%
\bibitem{BB84}
Bennett, C. H. \& Brassard, G. Quantum cryptography: public key distribution and coin tossing. \textit{In Proc. IEEE International Conference on Computers, Systems \& Signal Processing}, 175–179 (IEEE, NY, Bangalore, India, 1984).
%
\bibitem{nonlinear}
Curty, M., Jofre, M., Pruneri, V. \& Mitchell, M. W. Passive decoy-state quantum key distribution with coherent light. \textit{Entropy} \textbf{17}, 4064-4082 (2015).
%
\bibitem{nonlinear_2}
Boyd, R. W. Nonlinear optics. Academic press, Waltham, MA, USA (2008).
%
%
\bibitem{Mike}
Wang, W. \textit{et al}. Fully-passive quantum key distribution. \textit{arXiv preprint} arXiv:2207.05916 (2022).
%
\bibitem{Nielsen}
Nielsen, M. \& Chuang, I. \textit{Quantum Computation and Quantum Information}, Cambridge University Press (2000).
%
\bibitem{Renner1}
Renner, R. Symmetry of large physical systems implies independence of subsystems. \textit{Nat. Phys.} \textbf{3}, 645-649 (2007).
%
\bibitem{Renner2}
Christandl, M., König, R. \& Renner, R. Postselection technique for quantum channels with applications to quantum cryptography. \textit{Phys. Rev. Lett.} \textbf{102}, 020504 (2009).
%
\bibitem{uncertainty}
Tomamichel, M. \& Renner, R. Uncertainty relation for smooth entropies. \textit{Phys. Rev. Lett.} \textbf{106}, 110506 (2011).
%
\bibitem{Sun1}
Sun, S. H., Gao, M., Jiang, M. S., Li, C. Y. \& Liang, L. M. Partially random phase attack to the practical two-way quantum-key-distribution system. \textit{Phys. Rev. A} \textbf{85}, 032304 (2012).
%
\bibitem{Sun2}
Sun, S. H., Xu, F., Jiang, M. S., Ma, X. C., Lo, H. K. \& Liang, L. M. Effect of source tampering in the security of quantum cryptography. \textit{Phys. Rev. A} \textbf{92}, 022304 (2015).
%
\bibitem{Huang2}
Huang, A., Navarrete, Á., Sun, S. H., Chaiwongkhot, P., Curty, M. \& Makarov, V. Laser-seeding attack in quantum key distribution. \textit{Phys. Rev. Appl.} \textbf{12}, 064043 (2019).
%
\bibitem{Huang}
Huang, A., Li, R., Egorov, V., Tchouragoulov, S., Kumar, K. \& Makarov, V. Laser-damage attack against optical attenuators in quantum key distribution. \textit{Phys. Rev. Appl.} \textbf{13}, 034017 (2020).
%
\bibitem{Zhang}
Zhang, G., Primaatmaja, I. W., Haw, J. Y., Gong, X., Wang, C., \& Lim, C. C. W. Securing practical quantum communication systems with optical power limiters. \textit{PRX Quantum} \textbf{2}, 030304 (2021).
%
\bibitem{Posonova}
Ponosova, A., Ruzhitskaya, D., Chaiwongkhot, P., Egorov, V., Makarov, V. \& Huang, A. Protecting fiber-optic quantum key distribution sources against light-injection attacks. \textit{PRX Quantum} \textbf{3}, 040307 (2022).
%
\end{thebibliography}
\end{document}